\documentclass[]{aastex63}

\usepackage{amsmath}
\usepackage{amsfonts}
\usepackage{amssymb}

\usepackage{comment}

\usepackage{color}

\usepackage{tikz}

\newcommand{\diff}{\ensuremath{\mathrm{d}}}

\begin{document}

\title{
DAGnabbit! Ensuring Consistency between Noise and Detection in Hierarchical Bayesian Inference
}

\author{Reed Essick}
\email{essick@cita.utoronto.ca}
\affiliation{Canadian Institute for Theoretical Astrophysics, 60 St. George St, Toronto, Ontario M5S 3H8}
\affiliation{Department of Physics, University of Toronto, 60 St. George Street, Toronto, ON M5S 1A7}
\affiliation{David A. Dunlap Department of Astronomy, University of Toronto, 50 St. George Street, Toronto, ON M5S 3H4}

\author{Maya Fishbach}
\email{fishbach@cita.utoronto.ca}
\affiliation{Canadian Institute for Theoretical Astrophysics, 60 St. George St, Toronto, Ontario M5S 3H8}
\affiliation{Department of Physics, University of Toronto, 60 St. George Street, Toronto, ON M5S 1A7}
\affiliation{David A. Dunlap Department of Astronomy, University of Toronto, 50 St. George Street, Toronto, ON M5S 3H4}

\begin{abstract}
    Hierarchical Bayesian inference can simultaneously account for both measurement uncertainty and selection effects within astronomical catalogs.
    In particular, the hierarchy imposed encodes beliefs about the interdependence of the physical processes that generate the observed data.
    We show that several proposed approximations within the literature actually correspond to inferences that are incompatible with any physical detection process, which can be described by a directed acyclic graph (DAG).
    This generically leads to biases and is associated with the assumption that detectability is independent of the observed data given the true source parameters.
    We show several examples of how this error can affect astrophysical inferences based on catalogs of coalescing binaries observed through gravitational waves, including misestimating the redshift evolution of the merger rate as well as incorrectly inferring that General Relativity is the correct theory of gravity when it is not.
    In general, one cannot directly fit for the ``detected distribution'' and ``divide out'' the selection effects in post-processing. 
    Similarly, when comparing theoretical predictions to observations, it is better to simulate detected data (including both measurement noise and selection effects) rather than comparing estimates of the detected distributions of event parameters (which include only selection effects).
    While the biases introduced by model misspecification from incorrect assumptions may be smaller than statistical uncertainty for moderate catalog sizes (O(100) events), they will nevertheless pose a significant barrier to precision measurements of astrophysical populations.
\end{abstract}

%-------------------------------------------------

%-------------------------------------------------

\section{Introduction}
\label{sec:introduction}

Hierarchical Bayesian inference~\citep[e.g.,][]{Loredo:2004, Mandel:2009xr, Foreman_Mackey_2014, Mandel:2019} has become a cornerstone of analyses of astronomical catalogs as it provides a single formalism that can account for both measurement uncertainty and selection effects (i.e., the fact that some sources may be easier to detect than others).
Of particular interest is the application of hierarchical inference to learn about the properties of astrophysical populations of sources from a set of detected events (a.k.a. a catalog) and estimates of search sensitivity.
The goal of population inference is to measure a set of parameters that are common to all sources (``population parameters'') such as the maximum neutron star mass, the average black hole spin, or the correct theory of gravity.
Population parameters generally define a probability distribution from which each astrophysical source's individual parameters are drawn.
That is, the population parameters determine the distributions of the masses, spins, locations, and orientations that describe individual events.
Each event in turn gives rise to (noisy) data in a detector, which is a combination of the signal (a function of the source's true parameters) and the detector's noise (i.e. measurement error).
Finally, depending on the observed data (and potentially also the event's true parameters), there is some probability that determines whether the source actually makes it into the catalog (i.e., whether a search would detect this event).
These three-level hierarchical models (population parameters $\rightarrow$ source parameters $\rightarrow$ data and detection) are common across astrophysics.

Furthermore, such models can be expressed as directed acyclic graphs~\citep[DAGs;][]{Pearl:2009}, which provide a convenient shorthand for assumptions about the conditional (in)dependence between variates.
DAGs are generative models specifying our understanding of how the observed data come into being.
We must take care to construct inferences that are consistent with the known causal relations that govern how data are generated.

Although our conclusions apply to any probabilistic model consistent with Fig.~\ref{fig:general dag}, we focus on examples from gravitational-wave (GW) astronomy, specifically population inference of binary neutron stars (NSs) and black holes (BHs).
Since their first direct detection in 2015~\citep{GW150914}, GWs have provided a new and rapidly expanding view of the lives and deaths of massive stars~\citep{GWTC-1, GWTC-2, GWTC-2d1, GWTC-3}.
The latest catalog of GW transients detected by the advanced LIGO~\citep{LIGO}, advanced Virgo~\citep{Virgo}, and KAGRA~\citep{KAGRA} interferometers consists of $O(100)$ compact binary coalescences.
Studies of these events~\citep{GWTC-2-RnP, GWTC-3-RnP}, combined with other multi-messenger probes~\citep{GW170817-mma, GW170817-grb}, have greatly improved our understanding of the cosmos, but many open questions remain.
Catalogs of GW transients additionally present a relatively tractable application of hierarchical inference, as models of detector noise are generally reliable~\citep{Littenberg:2015, Biscoveanu:2020, Talbot:2020} and there are first-principles models of GW signals~\citep{Buonanno:1999, Bohe:2017, Nagar:2018, Ossokine:2020, Husa:2016, Khan:2016, Pratten:2020, Pratten:2021, Islam:2022, Varma:2019} along with the detector response~\citep{Cahillane:2017, Sun:2020}.
Semianalytic prescriptions~\citep{Essick:2023} can also be accurate approximations for the behavior of real searches~\citep[e.g.,][]{Messick:2017, Sachdev:2019, Hanna:2019, Cannon:2021, Allen:2004, DalCanton:2014, Usman:2015, Nitz:2017, Davies:2020, Allen:2005, Klimenko:2011, Klimenko:2004, Klimenko:2015, Adams:2015, Aubin:2020}.
As such, the entire data generation process can be directly simulated with high fidelity.
It is therefore of the utmost importance to guarantee that hierarchical analyses of GW catalogs are performed self-consistently to avoid spoiling the unique view through this new window on the universe.
We establish such consistency criteria.

We focus on two mistakes that commonly appear in the literature and show that they are inconsistent with physical models of how GW events are generated and detected. 
These common pitfalls are:
\begin{enumerate}
    \item \emph{Approximating detection probability as a function of source parameters rather than the observed data.}\\
        We show that the detection probability (``selection effects'') must be modeled consistently with the measurement error of each event's source parameters (``parameter estimation'').
        Thus, approximating the detection probability as a function of only the true source parameters generally leads to biased inference. 
        Furthermore, special care must be taken when generating mock catalogs to enforce this consistency.
    \item \emph{Directly fitting the distribution of detected events rather than astrophysical sources.}
        Specifically, we show that one cannot ``fit for the detected distribution'' and then later ``divide by selection effects'' without introducing biases in \emph{both} the detected \emph{and} astrophysical distributions. 
        Ignoring selection effects and fitting the detected distribution is therefore \emph{not} a viable alternative to correctly implementing the hierarchical Bayesian likelihood, even when one is only interested in the properties of the detected distribution.
\end{enumerate}
Fundamentally, both examples boil down to properly dealing with correlations induced within the data generation process by the fact that there is a single noise realization for any individual event.
That noise realization governs both parameter uncertainty and detectability.

We also remark that it is just as difficult to correctly predict the distribution of true parameters of detected events as it is to predict the astrophysical distribution or the distribution of detected data.
As such, to minimize the possibility of inconsistencies, comparisons between theory and observations should be carried out either between astrophysical distributions or in terms of the data recorded.
They should not be carried out over the latent single-event parameters.

The rest of this paper is structured as follows.
Sec.~\ref{sec:graphs} introduces the probabilistic models relevant to our hierarchical inference and demonstrates the implications of two different assumptions about the structure of conditional dependencies.
Sec.~\ref{sec:consistency mock pe with selection} then presents a concrete example of how bias can be introduced when parameter estimation and event selection are inconsistent.
Sec.~\ref{sec:detected distribution} takes this further and discusses why fitting for the ``detected distribution'' of true source parameters is inconsistent with physical models of how real data are generated.
We conclude in Sec.~\ref{sec:discussion}.

%-------------------------------------------------

\section{Probabilistic Graphs}
\label{sec:graphs}

We structure our discussion around probabilistic graphs, specifically DAGs~\citep{Pearl:2009}, that encode conditional dependencies (edges) between latent and observed variates (nodes).
As such, DAGs are shorthand for probabilistic models in which the distribution for any individual variate depends only on the other variates connected to it by inward pointing arrows.

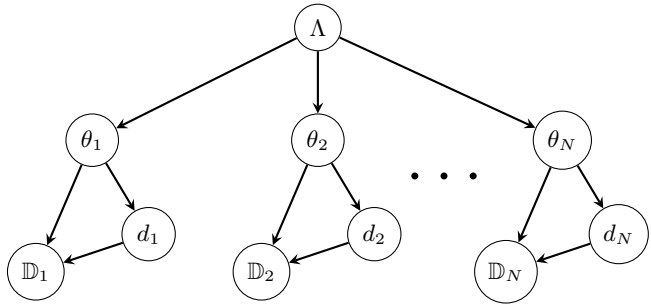
\begin{figure}
    \begin{center}
    \begin{tikzpicture}[node distance=1.5cm]
        \tikzstyle{block} = [circle, text centered, draw=black];
        \tikzstyle{arrow} = [thick, ->, >=stealth]
        \node (lambda) [block] {$\Lambda$};
        \node (theta1) [block, below of=lambda, xshift=-3.00cm] {$\theta_1$};
        \draw [arrow] (lambda) -- (theta1);
        \node (data1) [block, below of=theta1, xshift=+0.75cm, yshift=+0.25cm] {$d_1$};
        \draw [arrow] (theta1) -- (data1);
        \node (det1) [block, below of=theta1, xshift=-0.75cm, yshift=-0.25cm] {$\mathbb{D}_1$};
        \draw [arrow] (theta1) -- (det1);
        \draw [arrow] (data1) -- (det1);
        \node (theta2) [block, below of=lambda, xshift=+0.00cm] {$\theta_2$};
        \draw [arrow] (lambda) -- (theta2);
        \node (data2) [block, below of=theta2, xshift=+0.75cm, yshift=+0.25cm] {$d_2$};
        \draw [arrow] (theta2) -- (data2);
        \node (det2) [block, below of=theta2, xshift=-0.75cm, yshift=-0.25cm] {$\mathbb{D}_2$};
        \draw [arrow] (theta2) -- (det2);
        \draw [arrow] (data2) -- (det2);
        \node (dots) [below of=lambda, xshift=+1.75cm, yshift=-0.50cm] {\Huge{$\cdots$}};
        \node (thetaN) [block, below of=lambda, xshift=+3.25cm] {$\theta_N$};
        \draw [arrow] (lambda) -- (thetaN);
        \node (dataN) [block, below of=thetaN, xshift=+0.75cm, yshift=+0.25cm] {$d_N$};
        \draw [arrow] (thetaN) -- (dataN);
        \node (detN) [block, below of=thetaN, xshift=-0.75cm, yshift=-0.25cm] {$\mathbb{D}_N$};
        \draw [arrow] (thetaN) -- (detN);
        \draw [arrow] (dataN) -- (detN);
    \end{tikzpicture}
    \end{center}
    \caption{
        The most general DAG describing the process of constructing a catalog.
        Single-event parameters ($\theta_i$) for each event are drawn from the same population ($\Lambda$), and the data ($d_i$) and/or detection ($\mathbb{D}_i$) for each event depend only on the properties of that event.
        As we will show, common model misspecifications amount to different assumptions for what determines whether an event is detected (see Fig.~\ref{fig:simplified dags}).
    }
    \label{fig:general dag}
\end{figure}

Fig.~\ref{fig:general dag} shows the most general DAG we consider.
It relates the
\begin{itemize}
    \item [($\Lambda$)] population parameters (e.g., minimum and maximum mass, etc.), the
    \item [($\theta_i$)] single-event parameters for the $i^\mathrm{th}$ event (e.g., masses, spins, etc.), the
    \item [($d_i)$] $= n_i + h(\theta_i)$ observed data for the $i^\mathrm{th}$ event assuming additive noise $n_i$ and a signal model $h(\theta_i)$, and an
    \item [($\mathbb{D}_i$)] indicator signifying that the $i^\mathrm{th}$ event was detected.
\end{itemize}
We also assume that there is a single data set for each observation (e.g., GW strain timeseries) as opposed to multiple obseravtions with different instruments.\footnote{It is worth noting that other situations may involve more complicated DAGs, including marginalization over subsets of data observed in different instruments. See, e.g.,~\citet{Lieu:2017}.}
We further believe our data are generated by an inhomogeneous Poisson process~\citep[e.g.,][]{Loredo:2004, Mandel:2019}, which additionally relates the
\begin{itemize}
    \item [($\mathcal{K}$)] expected number of astrophysical events within the past light-cone spanning the duration of the experiment or, alternatively,
    \item [($K$)] $= \mathcal{K} P(\mathbb{D}|\Lambda)$ the expected number of detected events
\end{itemize}
where\footnote{Because individual events are either detected or not (a discrete set of possibilities), we write $P(\mathbb{D}|d,\theta)$ with a capital $P$ (probability mass function) to distinguish it from distributions over continuous variables, which we denote with lower-case $p$ (probability density function).}
\begin{equation}
    P(\mathbb{D}|\Lambda) = \int \diff\theta\, p(\theta|\Lambda) \int \diff d\, p(d|\theta) P(\mathbb{D}|d,\theta)
\end{equation}
is the probability that any individual event drawn from the astrophysical population would be detected.

Within this model, Fig.~\ref{fig:general dag} corresponds to the following joint distribution for a set of $N$ events.
%\begin{multline}\label{eq:general dag}
%    p(\{\theta_i, d_i, \mathbb{D}_i\}, N | \Lambda, \mathcal{K}) \propto \\
%        \mathcal{K}^N e^{-\mathcal{K} P(\mathbb{D}|\Lambda)} \prod\limits_i^N P(\mathbb{D}_i|d_i, \theta_i) p(d_i|\theta_i) p(\theta_i|\Lambda)
%\end{multline}
\begin{equation}\label{eq:general dag}
    p(\{\theta_i, d_i, \mathbb{D}_i\}, N | \Lambda, \mathcal{K}) \propto
        \mathcal{K}^N e^{-\mathcal{K} P(\mathbb{D}|\Lambda)} \prod\limits_i^N P(\mathbb{D}_i|d_i, \theta_i) p(d_i|\theta_i) p(\theta_i|\Lambda)
\end{equation}
where the single-event likelihood $p(d|\theta)$ is derived from the probability of observing a noise fluctuation equal to the residual between $d$ and $h(\theta)$.
\begin{equation}
    p(d|\theta) = p(n = d-h(\theta) | \theta)
\end{equation}
In general, the noise $n$ can depend on the source parameters $\theta$.
However, we typically assume that it does not or that it only depends on the time at which the event occurs (i.e., nonstationary noise).

Note also that $N$ is not Poisson distributed with mean $\mathcal{K}$, but it is Poisson distributed with mean $K$.
We will return to this point later.

\begin{figure}
    \begin{center}
    \begin{tikzpicture}[node distance=1.5cm]
        \tikzstyle{block} = [circle, text centered, draw=black];
        \tikzstyle{arrow} = [thick, ->, >=stealth]
        \node (right) {\Large{Physical}};
        \node (right model) [below of=right, yshift=+1.00cm] {$(\mathbb{D} \perp \theta \ | \ d)$};
        \node (wrong) [right of=right, xshift=+2.50cm] {\Large{Unphysical}};
        \node (wrong model) [below of=wrong, yshift=+1.00cm] {$(\mathbb{D} \perp d \ | \ \theta)$};
        \node (lambda right) [block, below of=right model, yshift=+0.50cm] {$\Lambda$};
        \node (theta1 right) [block, below of=lambda right] {$\theta$};
        \draw [arrow] (lambda right) -- (theta1 right);
        \node [below of=lambda right, xshift=+0.75cm, yshift=+0.75cm] {$p(\theta|\Lambda)$};
        \node (data1 right) [block, below of=theta1 right, xshift=+0.75cm, yshift=+0.25cm] {$d$};
        \draw [arrow] (theta1 right) -- (data1 right);
        \node [below of=theta1 right, xshift=+1.00cm, yshift=+1.10cm] {$p(d|\theta)$};
        \node (det1 right) [block, below of=theta1 right, xshift=-0.75cm, yshift=-0.25cm] {$\mathbb{D}$};
        \draw [arrow] (data1 right) -- (det1 right);
        \node [below of=theta1 right, xshift=+0.25cm, yshift=-0.40cm] {$P(\mathbb{D}|d)$};
        \node (lambda wrong) [block, below of=wrong model, yshift=+0.50cm] {$\Lambda$};
        \node (theta1 wrong) [block, below of=lambda wrong] {$\theta$};
        \draw [arrow] (lambda wrong) -- (theta1 wrong);
        \node [below of=lambda wrong, xshift=+0.75cm, yshift=+0.75cm] {$p(\theta|\Lambda)$};
        \node (data1 wrong) [block, below of=theta1 wrong, xshift=+0.75cm, yshift=+0.25cm] {$d$};
        \draw [arrow] (theta1 wrong) -- (data1 wrong);
        \node [below of=theta1 wrong, xshift=+1.00cm, yshift=+1.10cm] {$p(d|\theta)$};
        \node (det1 wrong) [block, below of=theta1 wrong, xshift=-0.75cm, yshift=-0.25cm] {$\mathbb{D}$};
        \draw [arrow] (theta1 wrong) -- (det1 wrong);
        \node [below of=theta1 wrong, xshift=-1.00cm, yshift=+0.75cm] {$Q(\mathbb{D}|\theta)$};
    \end{tikzpicture}
    \end{center}
    \caption{
        Two incompatible assumptions for the conditional dependencies within our hierarchical inference.
        Real experiments never have access to the true single-event parameters $\theta$, and therefore detection $\mathbb{D}$ can only depend on the observed data $d$ (\emph{left}).
        However, several proposed models incorrectly assume that the detection depends only on an event's true parameters (\emph{right}).
    }
    \label{fig:simplified dags}
\end{figure}
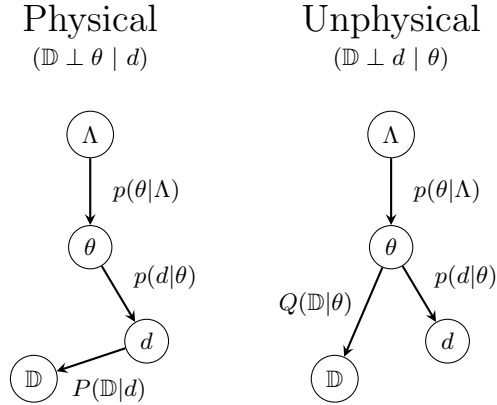

While Fig.~\ref{fig:general dag} is the most general DAG for signal detection, we do not actually believe that all the edges are physical.
That is, one often models reality with a simplified DAG.
Fig.~\ref{fig:simplified dags} shows two such options.
The left side ($\mathbb{D} \perp \theta \ | \ d$) assumes that detection depends only on the observed data.\footnote{Our notation $A \perp B \ | \ C$ means that $A$ and $B$ are conditionally independent given $C$. That is, $p(A,B|C) = p(A|C) p(B|C)$.}
Since real searches do not have access to $\theta$ directly, this must be the case for any physical detection process.
The right side ($\mathbb{D} \perp d \ | \ \theta$) shows a common approximation, that detection instead directly depends on the true source parameters and is independent of the data.

As we will show, these models are incompatible (assuming one model when the other is true causes biases) except when either
\begin{itemize}
    \item $d$ is one-to-one with $\theta$ (perfect measurement) or
    \item $\mathbb{D} \perp (d, \theta)$ : detection depends on neither $d$ nor $\theta$ (i.e., every event is equally detectable).
\end{itemize}
Neither condition holds in practice.
Measurements are noisy and some signals are easier to detect than others.
Hierarchical inference must therefore not only account for selection effects and measurement uncertainty but must take care to ensure that only self-consistent, physical assumptions about both are made.

Before presenting concrete examples of how models that assume ($\mathbb{D} \perp d \ | \ \theta$) introduce bias, we first examine the distributions implied by each DAG in Fig.~\ref{fig:simplified dags} in Secs.~\ref{sec:correct dag} and Sec.~\ref{sec:incorrect dag}.
In particular, we show that approaches claiming to fit the ``detected distribution'' implicitly assume ($\mathbb{D} \perp d \ | \ \theta$) in Sec.~\ref{sec:incorrect dag}.

%------------------------

\subsection{The Physical DAG $(\mathbb{D} \perp \theta \ | \ d)$}
\label{sec:correct dag}

Starting from Eq.~\ref{eq:general dag}, we make a few helpful manipulations.
First, we switch our parametrization from the expected number of astrophysical events $\mathcal{K}$ to the expected number of detections $K = \mathcal{K} P(\mathbb{D}|\Lambda)$, which allows us to write
%\begin{align}
%    p(\{\theta_i, & d_i, \mathbb{D}_i\}, \Lambda, K | N) \nonumber \\
%        & \propto p(\Lambda, K) \frac{K^N e^{-K}}{P(\mathbb{D}|\Lambda)^N} \prod\limits_i^N P(\mathbb{D}_i|d_i) p(d_i|\theta_i) p(\theta_i|\Lambda) \label{eq:general joint dist}
%\end{align}
\begin{equation}
    p(\{\theta_i, d_i, \mathbb{D}_i\}, \Lambda, K | N)
        \propto p(\Lambda, K) \frac{K^N e^{-K}}{P(\mathbb{D}|\Lambda)^N} \prod\limits_i^N P(\mathbb{D}_i|d_i) p(d_i|\theta_i) p(\theta_i|\Lambda) \label{eq:general joint dist}
\end{equation}
where we have included a prior over $(\Lambda, K)$ and used $(\mathbb{D} \perp \theta \ | \ d)$ to simplify $P(\mathbb{D}|d,\theta) = P(\mathbb{D}|d)$.
Note that this coordinate change completely separates the inference for $(K|N)$ from $(\{\theta_i, d_i, \mathbb{D}_i\}, \Lambda|N)$ assuming the prior $p(K,\Lambda)$ is separable.
This has previously been attributed to other mechanisms, such as marginalizing over the astrophysical rate 
with a specific prior $p(\mathcal{K}) \propto \mathcal{K}^{-1}$~\citep{Fishbach_2018, Essick:2023}.
However, marginalization is not strictly necessary; the appearance of the factor of $P(\mathbb{D}|\Lambda)^{-N}$ is generic.

Eq.~\ref{eq:general joint dist} appears frequently as a starting point within the literature~\citep[e.g.][]{GWTC-2-RnP, GWTC-3-RnP, Thrane:2019, Vitale:2020, Essick:2022a}.
Typically, though, this is not what is required for astrophysical inference.
Instead, we often wish to sample from a hyperposterior for $\Lambda$ conditioned on the observed data for each detected event and the fact that each event was detected.
\begin{widetext}
\begin{align}
    p(\Lambda | \{d_i, \mathbb{D}_i\}, N)
        & = \frac{\int \left[ \diff K \prod_i^N \diff\theta_i \right] p(\{\theta_i, d_i, \mathbb{D}_i\}, \Lambda, K | N)}{\int \left[ \diff\Lambda \diff K \prod_i^N \diff\theta_i \right] p(\{\theta_i, d_i, \mathbb{D}_i\}, \Lambda, K | N)} \nonumber \\
        & = \frac{\left(p(\Lambda) P(\mathbb{D}|\Lambda)^{-N} \prod_i^N P(\mathbb{D}_i|d_i) \int \diff\theta_i\, p(d_i|\theta_i) p(\theta_i|\Lambda) \right) \int \diff K\, p(N|K) p(K) }{\left(\int \diff\Lambda \left[ p(\Lambda) P(\mathbb{D}|\Lambda)^{-N} \prod_i^N P(\mathbb{D}_i|d_i) \int \diff\theta_i\, p(d_i|\theta_i) p(\theta_i|\Lambda) \right]\right) \int \diff K\, p(N|K) p(K) } \nonumber \\
        & \propto p(\Lambda) P(\mathbb{D}|\Lambda)^{-N} \prod\limits_i^N \int \diff\theta_i\, p(d_i|\theta_i)p(\theta_i|\Lambda) \label{eq:correct dag}
\end{align}
\end{widetext}
The integrals over $K$ in the numerator and the denominator cancel because they do not depend on any other variables.
This assumes a separable prior for $(K, \Lambda)$.
Note that the induced prior for $(\mathcal{K}, \Lambda)$ may not be separable.

In general, assumptions about $\mathcal{K}$ within a separable prior $p(\mathcal{K},\Lambda) = p(\mathcal{K})p(\Lambda)$ can still affect the inference of $\Lambda$.
However, assumptions about $K$ in a separable prior $p(K,\Lambda) = p(K)p(\Lambda)$ cannot.
The special case $p(\mathcal{K}) \sim 1/\mathcal{K}$ is the only prior on $\mathcal{K}$ that will not affect the inference for $\Lambda$ because it is scale invariant (i.e., it corresponds to $p(K) \sim 1/K$ independently of $\Lambda$).

There is also a term-by-term cancellation of $P(\mathbb{D}_i|d_i)$ in the numerator and denominator because we condition on both $d_i$ and $\mathbb{D}_i$.
That is, each $P(\mathbb{D}_i|d_i)$ factors out of the integrals over $\Lambda$ and $\{\theta_i\}$.
One can also (re)express Eq.~\ref{eq:correct dag} in terms of $p(d_i|\mathbb{D}_i,\Lambda) = p(d_i|\Lambda)P(\mathbb{D}|d_i)/P(\mathbb{D}|\Lambda)$ without introducing additional terms that depend on $\Lambda$, in which case the terms $P(\mathbb{D}|\Lambda)$ can be thought of as a normalization for the likelihood as suggested in~\citet{Mandel:2019}.
In this case, Eq.~\ref{eq:correct dag} becomes
\begin{equation}\label{eq:correct dag rearranged}
    p(\Lambda|\{d_i,\mathbb{D}_i\},N) \propto p(\Lambda) \prod_i^N p(d_i|\mathbb{D}_i,\Lambda).
\end{equation}
Combining this with the Poisson distribution for the observed number of events $P(N|K) \sim K^N e^{-K}$ is the ``bottom up'' derivation of the inhomogeneous Poisson likelihood suggested within~\citet{Mandel:2019}.

Within the literature, the absence of explicit $P(\mathbb{D}_i|d_i)$ terms in the final expression is often attributed to the assumption that detection is a deterministic function of the data (i.e., $P(\mathbb{D}_i|d_i) = 1$ always for detected data).
See, for example, discussion in~\citet{Essick:2022a}.
However, this cancellation is in fact more general and holds even if detection $\mathbb{D}$ is only a probabilistic function of the data $d$ as long as it is conditionally independent of the true source parameters $\theta$.

Eq.~\ref{eq:correct dag} is the workhorse behind most hierarchical Bayesian inferences based on GW catalogs.
See, e.g.,~\citet{Essick:2022b}, for discussion about common approximations for the high-dimensional integrals.

%------------------------

\subsection{The Unphysical DAG $(\mathbb{D} \perp d \ | \ \theta)$}
\label{sec:incorrect dag}

Following a similar approach with the assumption that $(\mathbb{D} \perp d \ | \ \theta$), we obtain\footnote{We denote probability density and mass functions that correspond to the physical detection model with $p$ and $P$. We denote the unphysical model with $q$ and $Q$.}
%\begin{multline}
%    q(\Lambda|\{d_i,\mathbb{D}_i\}, N) \propto \\
%        \frac{p(\Lambda)}{Q(\mathbb{D}|\Lambda)^N} \prod\limits_i^N \int \diff\theta_i\, p(d_i|\theta_i) p(\theta_i|\Lambda) Q(\mathbb{D}_i|\theta_i) \label{eq:incorrect dag}
%\end{multline}
\begin{equation}
    q(\Lambda|\{d_i,\mathbb{D}_i\}, N) \propto
        \frac{p(\Lambda)}{Q(\mathbb{D}|\Lambda)^N} \prod\limits_i^N \int \diff\theta_i\, p(d_i|\theta_i) p(\theta_i|\Lambda) Q(\mathbb{D}_i|\theta_i) \label{eq:incorrect dag}
\end{equation}
where
\begin{equation}
    Q(\mathbb{D}|\Lambda) = \int \diff\theta\, p(\theta|\Lambda) Q(\mathbb{D}|\theta) \label{eq:this equation}
\end{equation}
Note that there are factors of $Q(\mathbb{D}_i|\theta_i)$ within the numerator because these no longer factor out of the integrals over $\{\theta_i\}$ and therefore do not cancel between the numerator and denominator.
These can be thought of as a modification to the prior (see Appendix~\ref{sec:consistent pe and injections}).
However, just as we rearranged Eq.~\ref{eq:correct dag} to obtain Eq.~\ref{eq:correct dag rearranged}, it is instructive to further rearrange Eq.~\ref{eq:incorrect dag} as follows
\begin{align}
    q(\Lambda|\{d_i, \mathbb{D}_i\}, N)
        & \propto p(\Lambda) \prod\limits_i^N \int \diff\theta_i\, p(d_i|\theta_i) \frac{p(\theta_i|\Lambda) Q(\mathbb{D}_i|\theta_i)}{Q(\mathbb{D}|\Lambda)} \nonumber \\
        & = p(\Lambda) \prod\limits_i^N \int \diff\theta_i\, p(d_i|\theta_i) q(\theta_i|\mathbb{D}_i,\Lambda) \label{eq:incorrect dag rearranged}
\end{align}
where we recall the definition of $Q(\mathbb{D}|\Lambda)$ (Eq.~\ref{eq:this equation}), which acts as the normalization within $q(\theta|\mathbb{D},\Lambda)$.
Note that this rearrangement does not hold in more general DAGs (Fig.~\ref{fig:general dag}) and is uniquely associated with the assumption $(\mathbb{D} \perp d \ | \ \theta)$.
That is, if an inference is based on Eq.~\ref{eq:incorrect dag rearranged}, then it implicitly makes the unphysical assumption that $\mathbb{D} \perp d \ | \ \theta$.

Nevertheless, Eq.~\ref{eq:incorrect dag rearranged} has been proposed as an alternative to dealing with selection effects in several contexts~\citep{Isi:2019, Isi:2022, GWTC-2-TGR, GWTC-3-TGR, Talbot:2023, Sadiq:2022, Sadiq:2023, Rinaldi:2021}.
It is often referred to as fitting the ``detected distribution'' $q(\theta|\mathbb{D},\Lambda)$ rather than the ``astrophysical distribution'' $p(\theta|\Lambda)$.
However, we will show that Eq.~\ref{eq:incorrect dag rearranged} will generally \emph{not} yield a fit that is consistent with the true detected distribution (i.e., $q(\theta|\mathbb{D},\Lambda) \neq p(\theta|\mathbb{D},\Lambda) = p(\theta|\Lambda) \int \diff d\, P(\mathbb{D}|d) p(d|\theta) / P(\mathbb{D}|\Lambda)$), let alone lead to a consistent estimate for $p(\theta|\Lambda)$.
That being said, the severity of the bias can be problem-specific and may be small if the selection $P(\mathbb{D}|d, \theta)$ is only a slowly varying function.
See Sec.~\ref{sec:detected distribution} for more discussion.

%-------------------------------------------------

\begin{figure*}
    \begin{center}
        \includegraphics[width=0.85\textwidth]{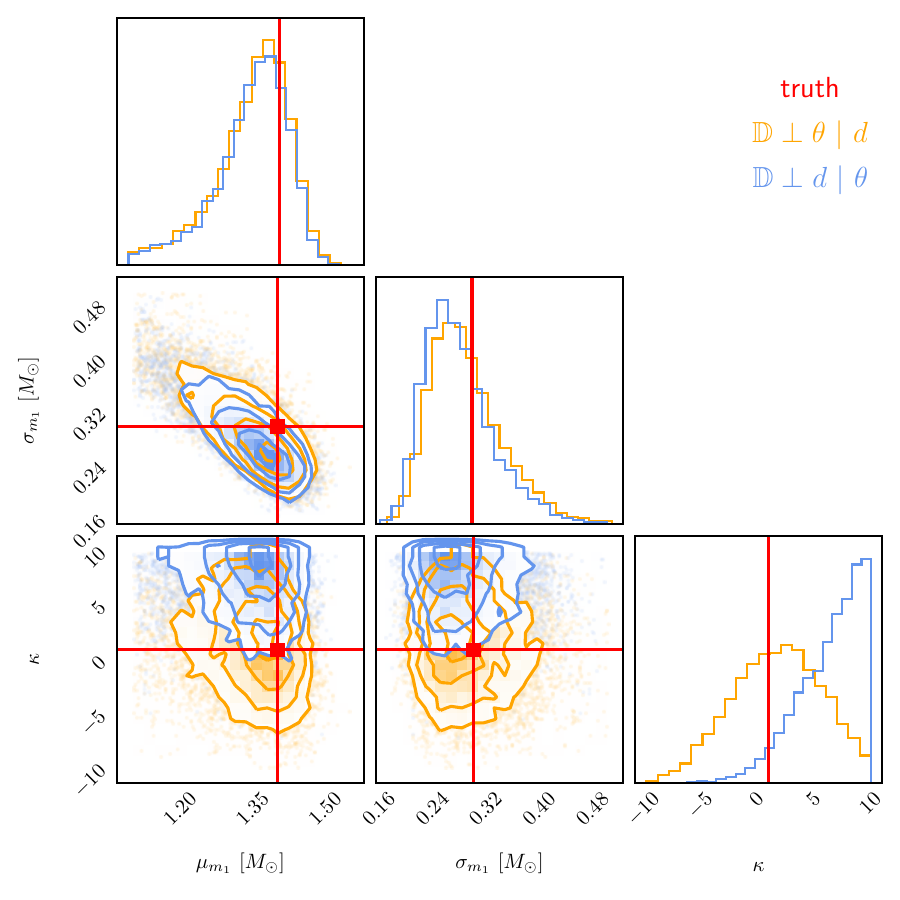}
    \end{center}
    \caption{
        Inferred hyperparameters with a catalog of 65 events with realistic measurement uncertainty using (\emph{orange}) the correct sensitivity estimate $P(\mathbb{D}|\rho_\mathrm{obs})$ and (\emph{blue}) the incorrect sensitivity estimate $Q(\mathbb{D}|\rho_\mathrm{opt})$. 
        Contours in the joint distributions enclose 0.5, 1, 1.5 and 2$\sigma$ of the posterior probability.
        While the inference of the mass distribution is not dramatically altered, we see that the redshift evolution $\kappa$ is inferred to be much larger than the real value when using the incorrect sensitivity estimate.
    }
    \label{fig:redshift evolution}
\end{figure*}

\section{Consistency between Parameter Estimation and Selection}
\label{sec:consistency mock pe with selection}

We first demonstrate a common mistake: using a physical model (Eq.~\ref{eq:correct dag}) to infer the astrophysical population but approximating the selection as a deterministic function of the true event parameters $\theta$.
This is the case when, for example, analysts use real parameter estimation samples based on $p(d|\theta)$ but estimate $Q(\mathbb{D}|\Lambda)$ semi-analytically with a threshold on the optimal signal-to-noise ratio ($\rho_\mathrm{opt}$, which is a function of the true event parameters; see Appendix~\ref{sec:consistent pe and injections} and~\citet{Essick:2023}) and use it in place of $P(\mathbb{D}|\Lambda)$ in Eq.~\ref{eq:correct dag}.
Specifically, as pointed out in \citet{Fishbach:2020}, data from systems at high redshift $z$ have some probability of being scattered by noise above the detection threshold.
As such, the posterior samples will correctly have support at high $z$.
However, if the selection is modeled as a deterministic function of the true parameters, it may have a sharp cutoff in $z$.
As such, the sensitivity at high $z$ can be dramatically underestimated, which then forces the inferred merger rate to be much larger in order to match the observed set of sources.\footnote{\citet{Essick:2023} described similar phenomena in the variance of $z$ in the distribution of detected sources $p(z|\mathbb{D},\Lambda)$. Models with selection based on the observed data have larger variances, and therefore support at higher $z$, than those based on $\rho_\mathrm{opt}$.}

This mistake can be thought of as incorrectly mixing different assumptions in different parts of the inference.
The posterior samples are generated assuming $(\mathbb{D} \perp \theta \ | \ d)$ but the sensitivity is estimated assuming $(\mathbb{D} \perp d \ | \ \theta)$.
Unsurprisingly, we find that one must make consistent assumptions throughout all aspects of the inference in order to avoid biases.
Indeed, a similar issue arises when analysts simulate mock events and corresponding sensitivity estimates with realistic detector noise (as specified, for example, by a power spectral density (PSD)), but generate \emph{different} noise realizations when creating mock parameter estimation samples and estimating the detector sensitivity.
Instead, the same noise realization must be used when generating parameter estimation samples and when measuring the selection effects in order to stay true to the physical DAG and avoid biases. 
We provide procedures to self-consistently generate both single-event posterior samples and sensitivity estimates in Appendix~\ref{sec:consistent pe and injections} based on~\citet{Fishbach:2020}; see also discussion in~\citet{Farah:2023} and the publicly available package \texttt{GWMockCat}~\citep{GWMockCat}.
Routines to self-consistently generate and select events based on the full data are also available in~\texttt{gw-distributions}~\citep{gw-distributions}.

%------------------------

\subsection{Example: Redshift Evolution of the Merger Rate}
\label{sec:redshift evolution}

We now present a concrete example of how misestimating $P(\mathbb{D}|\Lambda)$ by replacing it with $Q(\mathbb{D}|\Lambda)$ can bias an inference.

We generate a mock catalog of 65 binary NS systems drawn from an astrophysical mass and redshift distribution. 
We assume that the 
\begin{itemize}
    \item source-frame primary mass distribution is a truncated Gaussian between a minimum mass of $1\,M_\odot$ and a maximum mass of $2\,M_\odot$ with mean $\mu_{m_1}$ and standard deviation $\sigma_{m_1}$, the
    \item source-frame secondary masses are uniformly distributed between the minimum mass and the primary mass, and the
    \item merger rate (per unit of comoving volume and source-frame time) as a function of redshift follows a power law in $(1 + z)$ with index $\kappa$.
\end{itemize}
We further assume spins are zero and that sources are both uniformly distributed on the sky and isotropically distributed in orientations.
Once we have true source parameters drawn from an astrophysical distribution, we apply measurement uncertainty followed by a selection cut on the observed signal-to-noise ratio ($\rho_\mathrm{obs}$) as described in Appendix~\ref{sec:consistent pe and injections}.
To calculate $\rho_\mathrm{obs}$ and the corresponding selection cut, we assume a GW detector sensitivity given by the projected LIGO-A+ noise curve~\citep{AplusDesign}.
By following this procedure, we ensure that our mock GW data follow the generative model of Eq.~\ref{eq:correct dag}.

We then infer the population parameters ($\mu_{m_1}$, $\sigma_{m_1}$ and $\kappa$, fixing the other population parameters to their true values) with two different estimates for the detection probability: (physically with $P(\mathbb{D}|\Lambda)$) assuming that detection depends directly on the data through $\rho_\mathrm{obs}$ or (unphysically with $Q(\mathbb{D}|\Lambda)$) assuming that detection depends on the source parameters through $\rho_\mathrm{opt}$.
Fig.~\ref{fig:redshift evolution} shows the results.

As we can see, using a threshold on $\rho_\mathrm{opt}$ to model detectability instead of a threshold on $\rho_\mathrm{obs}$ significantly biases the inferred redshift evolution.
This is expected, as estimates of $Q(\mathbb{D}|\Lambda)$ based on $\rho_\mathrm{opt}$ will systematically underestimate the true detectability of systems $P(\mathbb{D}|\Lambda)$ at high redshift $z$.
The inference attempts to compensate by increasing the merger rate at high $z$, which it can only do by increasing $\kappa$.

We also note that the range of redshifts included within this study is relatively small ($z \lesssim 0.3$).
As such, this bias does not require large $z$ to become apparent and could be relevant for current GW catalogs~\citep{GWTC-1, GWTC-2, GWTC-2d1, GWTC-3}.

%-------------------------------------------------

\section{Inferring the Detected Distribution}
\label{sec:detected distribution}

We now turn our attention to another common approximation, namely fitting the ``detected distribution'' using Eq.~\ref{eq:incorrect dag rearranged}.
There are several reasons authors have adopted this approach, including uncertainty in how to estimate or incorporate selection effects~\citep{Rinaldi:2021, GWTC-2-TGR, GWTC-3-TGR, Isi:2019, Isi:2022}, concerns about the computational burden of Monte Carlo approximations~\citep{Talbot:2023}, or a desire to avoid implementing hierarchical inference altogether~\citep{Sadiq:2022, Sadiq:2023}.
As such, several authors claim that using Eq.~\ref{eq:incorrect dag rearranged} with a ``flexible enough'' model for $q(\theta|\mathbb{D},\Lambda)$ will allow them to recover the distribution of the true parameters of detected events.
When selection effects can be estimated, this is often extended to an inference for the astrophysical distribution by noting that
\begin{equation} \label{eq:divide by selection}
    p(\theta|\Lambda) = \frac{p(\theta|\mathbb{D},\Lambda)}{P(\mathbb{D}|\theta)} P(\mathbb{D}|\Lambda)
\end{equation}
within the physical DAG and assuming that $q(\theta|\mathbb{D},\Lambda)$ is an accurate approximation for $p(\theta|\mathbb{D},\Lambda)$.
That is, they estimate
\begin{equation}
    q(\theta|\Lambda) \propto \frac{q(\theta|\mathbb{D},\Lambda)}{P(\mathbb{D}|\theta)} \ .
\end{equation}
However, Eq.~\ref{eq:incorrect dag rearranged} is incompatible with $\mathbb{D} \perp \theta \ | \ d$ (Eq.~\ref{eq:correct dag}).
As such, it will not lead to the correct $p(\theta|\mathbb{D},\Lambda)$ (i.e., $q(\theta|\mathbb{D},\Lambda) \neq p(\theta|\mathbb{D},\Lambda)$ in general) or the correct astrophysical distribution ($q(\theta|\Lambda) \neq p(\theta|\Lambda)$).
We describe why this is the case below.

Inferences based on Eq.~\ref{eq:incorrect dag rearranged} will tend to produce distributions $q(\theta|\mathbb{D},\Lambda)$ that are too tight (smaller variance) than the true distribution $p(\theta|\mathbb{D},\Lambda)$.
We can see this by noting that the noise is additive and (often) independent of $\theta$.
As such, we can write the variance of the data with respect to the distribution $p(n,\theta)$ as
\begin{equation}
    \mathrm{V}[d]_{p(n,\theta)} = \mathrm{V}[n]_{p(n)} + \mathrm{V}[h(\theta)]_{p(\theta)}
\end{equation}
which makes it clear that, in the absence of selection effects, the variance of the data must be larger than the variance from uncertainty in $\theta$ alone because $\mathrm{V}[n]_{p(n)} > 0$.
Indeed, Eq.~\ref{eq:incorrect dag rearranged} forces the inferred $q(\theta|\mathbb{D}, \Lambda)$ to be narrow as it will try to match the predicted variance in $d$ to what is contained within the observed catalog.

On the other hand, in the presence of selection effects, there is some non-zero probability that $d$ will be scattered by noise into the detectable region of data-space for any $\theta$.
This implies that the true $p(\theta|\mathbb{D},\Lambda)$ should be wider (larger variance) than the detected distribution of data $p(d|\mathbb{D},\Lambda)$.

Eq.~\ref{eq:incorrect dag rearranged} therefore will lead to an inferred $q(\theta|\mathbb{D},\Lambda)$ that is narrower than what would be inferred via Eq.~\ref{eq:correct dag} (i.e., the true distribution $p(\theta|\mathbb{D},\Lambda)$).
Both approaches attempt to match the observed variance in $d$, but the differences in their assumptions for how data is scattered and events are detected lead to opposite effects for the variance of $(\theta|\mathbb{D},\Lambda)$.
To put a finer point on this, Eq.~\ref{eq:incorrect dag rearranged} will produce a biased estimate ($q(\theta|\mathbb{D},\Lambda) \neq p(\theta|\mathbb{D},\Lambda)$) if the data are generated by any physical process subject to nontrivial measurement uncertainty and selection effects even though Eq.~\ref{eq:incorrect dag rearranged} may still be able to perfectly match the true distribution of detected data ($q(d|\mathbb{D},\Lambda) = p(d|\mathbb{D},\Lambda)$).

%------------------------

\subsection{Example: BBH population inference}
\label{sec:neutron star masses}

\begin{figure*}
    \begin{center}
        \includegraphics[width=1.0\textwidth, clip=True, trim=0.0cm 0.30cm 0.0cm 0.0cm]{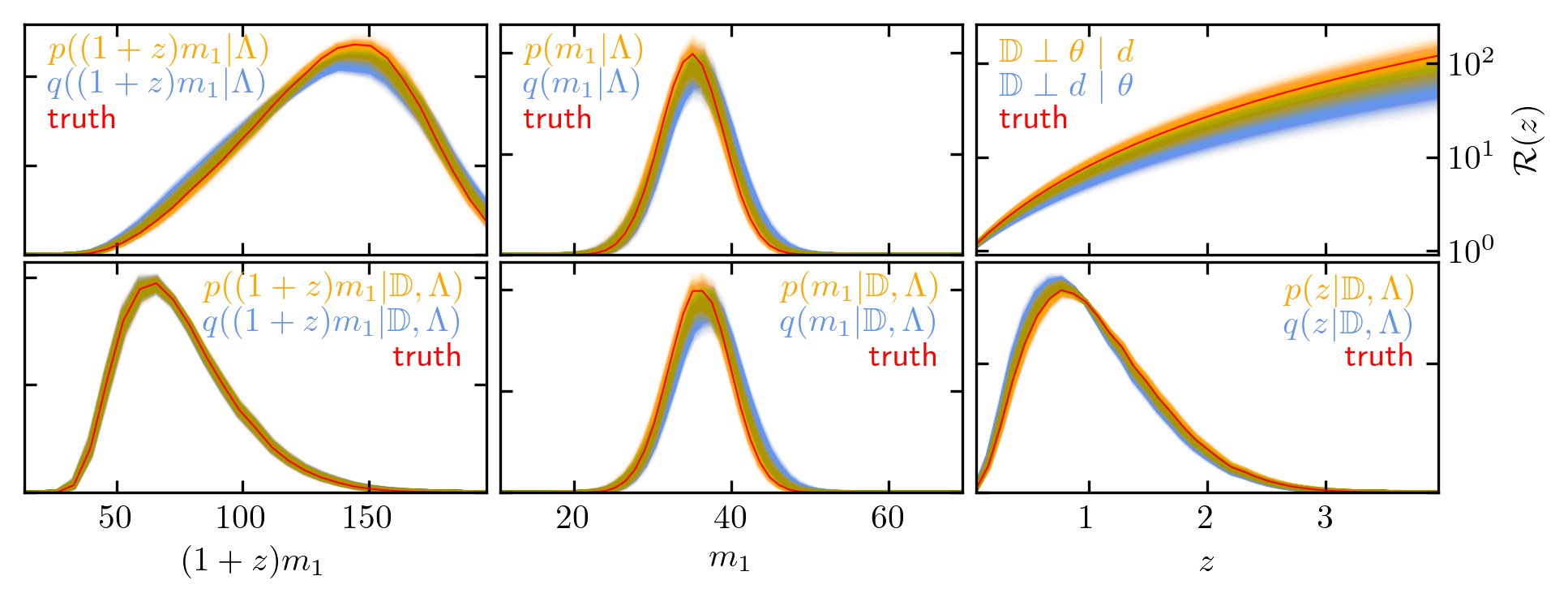}
    \end{center}
    \caption{
        Detector-frame (i.e., redshifted) mass (\emph{left}), source-frame mass (\emph{middle}) and redshift (\emph{right}) distributions inferred with 805 mock events at LIGO-A+ sensitivity.
        (\emph{top}) The astrophysical distributions and (\emph{bottom}) detected distributions with (\emph{orange}) draws from the hyperposterior inferred under the physical DAG, which correctly recover (\emph{red}) the true injected population, and (\emph{blue}) draws from the hyperposterior inferred under the unphysical DAG, which, especially for the redshift distribution, fails to recover either the true astrophysical or the true detected distribution.
    }
    \label{fig:BBH_mass_redshift}
\end{figure*}

We now consider what happens when we try to infer the merger rate as a function of masses and redshift with the unphysical DAG.
We simulate a population of GW observations drawn from an underlying astrophysical distribution, applying measurement uncertainty consistently with selection effects in accordance with the astrophysical DAG using the procedure described in Appendix~\ref{sec:consistent pe and injections}.
For simplicity, we assume a similar model to Sec.~\ref{sec:consistency mock pe with selection}:
\begin{itemize}
    \item source-frame primary masses $m_1$ are drawn from a Gaussian distribution centered at $35\,M_\odot$ with a standard deviation of $4\,M_\odot$,
    \item source-frame secondary masses $m_2$ follow a flat distribution between $1\,M_\odot$ and $m_1$, and the 
    \item merger rate increases with redshift proportionally to $(1 + z)^3$ out to a horizon resdhift of $z  = 4$.
\end{itemize}
To determine which sources are observed, we again use a representative LIGO-A+ sensitivity~\citep{AplusDesign}.
We generate 805 detected events from this population.

We carry out a population inference of this mock catalog using both the physical and unphysical DAGs in turn.
We fit the primary mass distribution and the rate as a function of redshift, fixing the secondary mass distribution (equivalently, the mass ratio distribution) to the true distribution.
We again assume that all binaries are nonspinning.

Fig.~\ref{fig:BBH_mass_redshift} shows the results in terms of the recovered redshifted primary mass, source-frame primary mass, and redshift distributions.
Unsurprisingly, when we use the likelihood given by the physical DAG, we recover the correct mass and redshift distributions. 
However, when we attempt to fit for the detected population using the unphysical DAG, we generally recover biased results, even though we assume a functional form that can correctly describe the true distributions (i.e., it is flexible enough to support the correct distribution).
The bias in the recovered redshifted mass distribution is small in this example because redshifted masses are relatively well measured (particularly because we fix the mass ratio distribution).
The bias is greatest in the redshift distribution because GW selection effects are a strong function of redshift, and the redshift is poorly-measured for individual events.
In other words, if we think of detectability as a function of true source parameters, as in the unphysical DAG, an event's detectability as a function of merger redshift varies significantly within its typical measurement uncertainty.
This leads the unphysical DAG to recover a detected redshift distribution that is too narrow compared to reality (i.e. has less support at high redshifts). 

We therefore expect any GW population inference based on the unphysical DAG to fail most noticeably when it comes to measuring evolution with redshift. 
Because of correlations between the source-frame mass distribution and merger redshift, this bias in the redshift distribution also impacts the source-frame mass distribution, which, when inferred under the unphysical DAG, excludes the correct mass distribution at $>3\sigma$ in our example.

With small catalogs that have large statistical uncertainties, the bias in the mass distribution recovered with the unphysical DAG may be small by eye, which is why current fits to the ``detected mass distribution" agree (within statistical uncertainty) with results using the physical DAG~\citep{Rinaldi:2021, Sadiq:2022, Sadiq:2023}, especially when they fix the redshift distribution. 
Nevertheless, precisely measuring these features is key to unlocking the origins of BHs and NSs, using them for GW cosmology, and understanding the composition of NSs.
This can only be done when using the correct likelihood.

We strongly caution against using the unphysical DAG to make any precise statements about astrophysical populations, such as the locations and widths of features in the mass or spin distribution, or the evolution of these features with redshift, which is especially prone to bias.

%------------------------

\subsection{Example: Deviations from General Relativity}
\label{sec:tgr}

\begin{figure*}
    \begin{minipage}{0.49\textwidth}
        \begin{center}
            {\Large wide population ($\sigma_\Lambda = 3$)} \\
            \includegraphics[width=1.0\columnwidth]{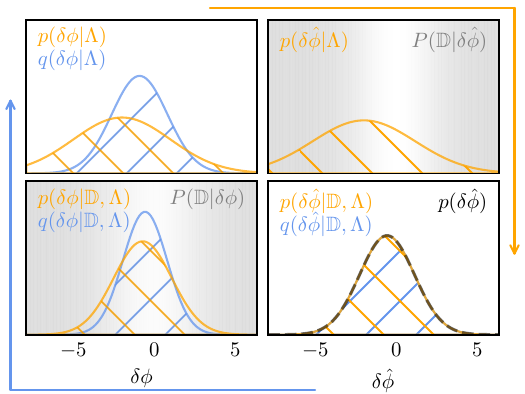}
        \end{center}
    \end{minipage}
    \begin{minipage}{0.49\textwidth}
        \begin{center}
            {\Large narrow population ($\sigma_\Lambda \approx 0.6$)} \\ % stdv_pop = 0.5775
            \includegraphics[width=1.0\columnwidth]{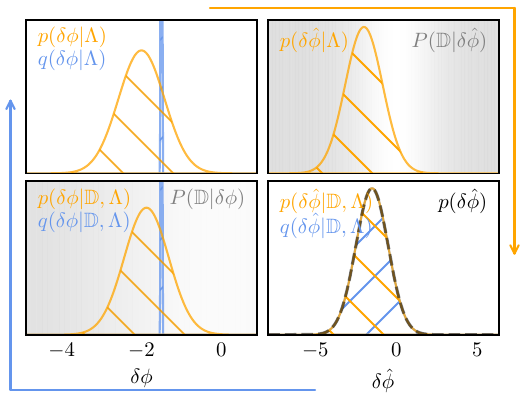}
        \end{center}
    \end{minipage}
    \caption{
        Distributions inferred in the limit of an infinite number of events within our Gaussian model of a hierarchical test of GR with both the (\emph{orange}, Eq.~\ref{eq:correct dag}) physical DAG and (\emph{blue}, Eq.~\ref{eq:incorrect dag rearranged}) unphysical DAG.
        We assume $\mu_\Lambda = -2$, $\sigma_o = 1$, $\mu_D = 0$, and $\sigma_D = 2$.
        Each group of panels has exactly the same parameters except for the width of the true astrophysical population, which is either (\emph{left panels}) wide ($\sigma_\Lambda = 3$) or (\emph{right panels}) narrow ($\sigma_\Lambda \approx 0.6$) compared to the measurement uncertainty.
        Each set of panels shows the (\emph{top left}) inferred astrophysical distributions of the deviation parameter $\delta\phi$, (\emph{top right}) inferred distribution of maximum-likelihood estimates for the deviation parameter $\delta\hat\phi$ before conditioning on detection under the physical DAG; shading denotes the probability of detection based on the data $P(\mathbb{D}|\delta\hat\phi)$, (\emph{bottom left}) inferred distributions of $\delta\phi$ for detected events; shading denotes the probability of detection based on the deviation parameter under the physical DAG $P(\mathbb{D}|\delta\phi)$, which is used to compute $q(\delta\phi|\Lambda) = q(\delta\phi|\mathbb{D},\Lambda) / P(\mathbb{D}|\delta\phi)$, and the (\emph{bottom right}) observed distribution of $\delta\hat\phi$ for each inference and (\emph{black dashed line}) the actual observed distribution (i.e., data).
        Arrows around the edges of the panels show the direction followed by each inference to connect the astrophysical distribution (\emph{top left}) and the observed distribution of data (\emph{bottom right}).
    }
    \label{fig:tgr}
\end{figure*}

As a final worked example, we consider an analytic model for parametrized deviations from General Relativity (GR).
Specifically, we consider a single deviation parameter ($\delta\phi$) inspired by the post-Einsteinian parametrization~\citep{Cornish:2011} as implemented within current LIGO-Virgo-KAGRA analyses~\citep{Agathos:2014, GWTC-2-TGR, GWTC-3-TGR}.
Such terms capture differences in the evolution of a GW's phase with frequency and are defined such that GR is recovered in the limit $\delta \phi \rightarrow 0$.
It can be difficult to precisely quantify the sensitivity of existing searches (which implicitly assume GR is correct) to arbitrary non-GR signals.\footnote{See, e.g.,~\citet{Narola:2023} for an example for sensitivity estimates from a single search assuming a single parametrization for deviations as well as~\citet{Chia:2020} and~\citet{Chia:2023} for other templated searches.~\citet{Payne:2023} presents a hierarchical inference that includes selection effects based on, e.g., the mass and distance but neglects the selection over deviations from GR.}
Nevertheless, we may generally expect that the probability of detection will be lower for data that seems to contain larger deviations.
We implement a simple model using this intuition and show that Eq.~\ref{eq:incorrect dag rearranged} leads to biases.

To wit, we consider a hypothetical experiment that can constrain $\delta\phi$ using a catalog of events.
Each event will have its own true $\delta\phi_i$ (i.e., $\theta$) and corresponding maximum-likelihood estimate $\delta\hat\phi_i$ (i.e., $d$).
We then model the experiment as a combination of Gaussians.
That is, 
\begin{itemize}
    \item $\delta\phi_i$ is Gaussian distributed according to a population model with unknown mean ($\mu_\Lambda$) and variance ($\sigma_\Lambda^2$),
    \item $\delta\hat\phi_i$ is Gaussian distributed about the true $\delta\phi_i$ with known variance ($\sigma_o^2$), and
    \item a system is detected with Gaussian probability as a function of $\delta\hat\phi$ centered on $\mu_D = 0$ with a known variance ($\sigma_D^2$).
\end{itemize}
See Appendix~\ref{sec:gaussian toy models} for explicit expressions.
Because all the relevant integrals are Gaussian, we consider a Gaussian ansatz for $q(\delta\phi|\mathbb{D},\Lambda)$ without fear of introducing additional model systematics.
We then examine the behavior of inferences based on Eq.~\ref{eq:correct dag} and Eq.~\ref{eq:incorrect dag rearranged} analytically.

Fig.~\ref{fig:tgr} shows our main conclusions in the limit of an infinite number of events.
Again, see Appendix~\ref{sec:gaussian toy models} for explicit expressions.
Briefly, the inference based on Eq.~\ref{eq:correct dag} correctly recovers all the relevant distributions: $p(\theta|\Lambda)$, $p(\theta|\mathbb{D},\Lambda)$, and $p(d|\mathbb{D},\Lambda)$.
However, even though an inference based on Eq.~\ref{eq:incorrect dag rearranged} may be able to recover $q(d|\mathbb{D},\Lambda) = p(d|\mathbb{D}, \Lambda)$, we see that the other inferred distributions are incorrect: $q(\theta|\mathbb{D},\Lambda) \neq p(\theta|\mathbb{D},\Lambda)$ and $q(\theta|\Lambda) \neq p(\theta|\Lambda)$.
In fact, Appendix~\ref{sec:gaussian toy models} demonstrates that Eq.~\ref{eq:incorrect dag rearranged} may not even be able to correctly recover $q(d|\mathbb{D},\Lambda) = p(d|\mathbb{D},\Lambda)$.
It additionally considers a model with deterministic (rather than probabilistic) detection $P(\mathbb{D}|d)$, finding similar behavior.

While these conclusions hold quite generally, it is interesting to consider the limit in which the population is very narrow (i.e., all events have the similar $\delta\phi$).
This is shown in the right-hand panels in Fig.~\ref{fig:tgr}.
In this case, the unphysical DAG (Eq.~\ref{eq:incorrect dag rearranged}) will infer $q(\delta\phi|\mathbb{D},\Lambda) \rightarrow \delta(\delta\phi - m_{\delta\hat\phi})$, where $m_{\delta\hat\phi}$ is the mean of the observed maximum-likelihood estimates.
In general, however, this will \emph{not} correspond to the true mean of the population; it will be shifted away from the true mean and towards the center of the detection window.
To put this another way, an inference based on Eq.~\ref{eq:incorrect dag rearranged} could confidently (correctly) infer that GR is not the correct theory of gravity when that is true.
However, in that case, it could equally confidently (incorrectly) infer the wrong value for $\delta\phi$ for all events and therefore the wrong theory of gravity.

To summarize, ``fitting the detected distribution'' via Eq.~\ref{eq:incorrect dag rearranged} can lead to constraints that are artificially tight and centered on the incorrect value, even at the level of the distribution of $\delta\phi$ for detected events: $q(\delta\phi|\mathbb{D},\Lambda)$.
This, in turn, could lead one to conclude that GR is the correct theory of gravity even when it is not or to infer the incorrect value for a deviation from GR.

%-------------------------------------------------

\section{Discussion}
\label{sec:discussion}

We have shown that inconsistencies in generative models for observed data can lead to significant biases in the resulting inferences.
In particular, we stress that no physical detection process will have access to the true parameters of any event and, as such, detection can only depend on the observed data.
The consequences of this simple idea are far reaching, and we show that several approaches commonly used within the literature are inconsistent with physical detection processes.
Such inferences are therefore intrinsically biased.

Perhaps the most profound is the observation that attempts to infer the ``detected distribution'' based on Eq.~\ref{eq:incorrect dag rearranged} implicitly assumes $\mathbb{D} \perp d \ | \ \theta$.
That is, they assume that detection is independent of the observed data.
This causes them to generically infer $q(\theta|\mathbb{D},\Lambda)$ that are too narrow and could lead to posterior credible regions that do not contain true distribution.
This has immediate consequences for proposed inferences of the mass, spin, redshift distributions of merging compact binaries as well as tests of GR based on catalogs of GW transients.

Importantly, we find that biases from assuming $p(\theta|\mathbb{D},\Lambda) = q(\theta|\mathbb{D},\Lambda)$ inferred from Eq.~\ref{eq:incorrect dag rearranged} remain even if the model can exactly match the true distribution of detected data $p(d|\mathbb{D},\Lambda)$.
As such, the data alone may not always tell us whether the assumptions behind Eq.~\ref{eq:incorrect dag rearranged} are unphysical.
One should instead take care to guarantee that the prior assumptions about causal structure implicit within hierarchical inference always reflect our understanding of physical processes.

That being said, there are conditions in which the bias introduced may be small.
These are when the assumptions behind the two DAGs in Fig.~\ref{fig:simplified dags} are compatible.
Specifically, when measurement uncertainty is small ($d$ is one-to-one with $\theta$) or when all events are equally detectable ($\mathbb{D}$ does not depend on either $d$ or $\theta$).
The latter approximately applies if measurements are precise enough that the data are clustered within a region over which $P(\mathbb{D}|d)$ does not change significantly.
See Appendix~\ref{sec:gaussian toy models} for explicit examples in the context of Gaussian models.
That may be the case with some tests of GR, but is unlikely to be the case for inferences that depend on the redshifts of GW coalescences.

As a corollary of our results, we argue against performing model comparison on the ``detected distribution" $p(\theta|\mathbb{D},\Lambda)$.
Often, theorists will simulate an astrophysical population $p(\theta|\Lambda)$ (i.e. population synthesis) and wish to compare it to observations.
It is common practice to then apply some model for $P(\mathbb{D}|\theta)$ and generate a prediction for $p(\theta|\mathbb{D},\Lambda)$.
Even though this detected distribution may be physically meaningful, it is unnecessary and often misleading.
From the data, it is just as (if not more) straightforward to infer the astrophysical population $p(\theta | \Lambda)$ than the detected distribution $p(\theta|\mathbb{D},\Lambda)$; there is no shortcut that gives the detected distribution more readily than the astrophysical one.
While it is sometimes possible to eye-ball whether individual events' parameter estimation samples are roughly consistent with a predicted ``detected distribution,'' such comparisons are not statistically meaningful.
As such, they do not provide a principled alternative to hierarchical Bayesian inference for a variety of reasons~\citep[e.g., sensitivity to reference parameter estimation priors;][]{Fishbach:2020, 2020PhRvD.102h3026G, 2020ApJ...895..128M, Essick:2021, Moore:2021, 2023MNRAS.525.3986M}.
If one wishes to avoid fitting the true astrophysical distribution, a valid option is to instead fit the distribution of detected \emph{data}.
One can then perform model comparison by forward-modeling the detected data.
In other words, simulation-based inference~\citep[for a review, see, e.g.,][]{2020PNAS..11730055C} is a valid alternative to evaluating the hierarchical Bayesian likelihood, but models must be evaluated based on their predicted data rather than the predicted true parameters.

Finally, we review a flexible, light-weight algorithm to generate self-consistent single-event parameter uncertainties and sensitivity estimates based on the data for synthetic catalogs in Appendix~\ref{sec:consistent pe and injections}.
Any forecast for population inference with current or proposed GW detectors~\citep{NextGeneration, CosmicExplorer, EinsteinTelescope} should take care to implement such procedures in order to avoid the current pitfalls that plague many existing publications that use mock catalogs.

%-------------------------------------------------

\acknowledgments

We sincerely thank Amanda Farah and Daniel Holz for helpful discussions at early stages of this project as well as Max Isi and Will Farr for their insightful suggestions.
We are also thankful to the organizers of GWPopNext (University of of Milano-Bicocca, July 2023), especially lead organizer Davide Gerosa, for providing a venue where we refined many of these ideas.

R.E. and M.F. are supported by the Natural Sciences \& Engineering Research Council of Canada (NSERC).

This work used the following software: \texttt{numpy}~\citep{numpy}, \texttt{scipy}~\citep{scipy}, \texttt{matplotlib}~\citep{matplotlib}, \texttt{jax}~\citep{jax}, \texttt{numpyro}~\citep{numpyro}, \texttt{pyro}~\citep{pyro}, \texttt{lalsuite}~\citep{lalsuite}, and \texttt{h5py}~\citep{h5py}.

%-------------------------------------------------

\newpage

\appendix

%\twocolumngrid

%-------------------------------------------------

\section{Self-consistent procedures to generate single-event posterior samples and sets of detected injections}
\label{sec:consistent pe and injections}

It is often useful to simulate synthetic observations in order to test analysis techniques in a controlled setting.
We describe consistency requirements for generating mock catalogs of GW events: the measurement uncertainty applied to the parameters of individual events must be consistent with the assumed selection function.
We distinguish between selection functions based on sources' true parameters $\theta$ and selection functions based only on the observed data $d$.
We also clarify what these different choices imply for single-event parameter estimation.

Within GW catalogs, individual events' parameters are typically estimated separately for each event with a reference set of prior assumptions, yielding reference single-event posteriors.
These encode the likelihood $p(d|\theta)$ in a more convenient representation~\citep[see discussion in ][]{Essick:2022b}.
Meanwhile, selection effects are measured by Monte Carlo integrals that approximate search sensitivity to different sources, usually by injecting simulated signals into the real detector data and running real searches to determine which are detected~\citep{GWTC-3-injections}.
In order to simulate this process, we must generate a synthetic catalog with self-consistent assumptions about the individual-event measurement uncertainty and the selection criteria.

It is common to model the detection probability by a deterministic threshold on some statistic $\mathbb{S}$ of the observed data and/or the event's parameters.
That is,
\begin{equation}
    P(\mathbb{D}|d, \theta) = \Theta( \mathbb{S}(d, \theta) \geq \mathbb{S}_\mathrm{thr})
\end{equation}
where $\Theta$ is the indicator function.
Again, because we do not observe the events' true parameters $\theta$ in real catalogs, the real selection must be a function of only the observed data $d$.
However, it is also common to instead model selection as a threshold on the event's true parameters, which is usually based on the optimal signal-to-noise ratio ($\rho_\mathrm{opt}$)
\begin{equation}
\label{eq:rho-opt}
    \rho^2_\mathrm{opt}(\theta) \equiv 4\int \diff f\, \frac{|h(\theta)|^2}{S}
\end{equation}
for a signal model $h(\theta)$ and one-sided power spectral density $S$.
See~\citet{Essick:2023} for an extensive discussion about the differences between $\rho_\mathrm{opt}$ and observed detection statistics.

Note that some authors (ourselves included) use the notation $P(\mathbb{D}|\theta)$ to mean
\begin{equation}
    P(\mathbb{D}|\theta) = \int \diff d\, P(\mathbb{D} | d) p(d | \theta) \label{eq:this one right here}
\end{equation}
This should not be confused with the case where the selection function directly depends on $\theta$, which we instead denote by $Q(\mathbb{D}|\theta)$.
We will always distinguish between the different assumptions for the selection function by writing the functional dependence of the detection statistic: either $\mathbb{S}(d)$ or $\mathbb{S}(\theta)$.

%------------------------

\subsection{Selection Based on Observed Data: $\mathbb{D} \perp \theta \ | \ d$}
\label{sec:data}

The most realistic approach models event selection as a deterministic function of (statistics derived from) the observed data.
\begin{equation}
    \mathbb{S}(d, \theta) = \mathbb{S}(d)
\end{equation}
With this approach, we know that $P(\mathbb{D}|d, \theta) = \Theta(\mathbb{S}(d)\geq \mathbb{S}_\mathrm{thr}) = 1$ axiomatically for all detected data.

The most straightforward way to simulate a catalog is simulate the full data.
That is, draw single-event parameters $\theta \sim p(\theta|\Lambda)$, then draw data $d \sim p(d|\theta)$, and finally determine whether the event is detected based on $\mathbb{S}(d)$ and $\mathbb{S}_\mathrm{thr}$.
However, approximations are also used.
Specifically, it is common to assume that search sensitivity is driven primarily by the observed signal-to-noise ratio ($\rho_\mathrm{obs}$).
This statistic follows a known distribution under stationary Gaussian noise centered on $\rho_\mathrm{opt}$.
Therefore, sensitivity estimates can be created by simulating $\rho_\mathrm{obs}$ from $\rho_\mathrm{opt}$ without generating the actual data.
Again, see~\citet{Essick:2023} for more details.

Either way, one must generate single-event posterior samples that are consistent with the selection procedure.
If the full data are simulated and recorded, then it is straightforward to sample from the posterior conditioned on that (axiomatically detectable) data~\citep[see, e.g.,][]{Veitch:2015, Ashton:2019, Ashton:2021}.
However, if only $\rho_\mathrm{obs}$ is simulated, then one must take care to simulate uncertainty in the posterior that is consistent with the noise fluctuations implicitly included in $\rho_\mathrm{obs}$.
Procedures for this were presented in~\citet{Fishbach:2020} and~\citet{Farah:2023}.
We provide a similar prescription below.

Fundamentally, the issue is that we have assumed a particular functional form for $p(\rho_\mathrm{obs}|\theta)$ which encodes noise fluctuations within the selection procedure.
The rest of the likelihood model must respect this.
Normally, one might write the distribution over both the data and $\rho_\mathrm{obs}$ as
\begin{equation}
    p(d, \rho_\mathrm{obs}|\theta) = p(d|\theta) p(\rho_\mathrm{obs}|d)
\end{equation}
where $p(\rho_\mathrm{obs}|d)$ is deterministic.
However, we can instead factor the distribution as
\begin{align}
    p(d, \rho_\mathrm{obs}|\theta)
        & = p(d|\rho_\mathrm{obs},\theta) p(\rho_\mathrm{obs}|\theta) \nonumber \\
        & = p(d|\rho_\mathrm{obs}, \theta) p(\rho_\mathrm{obs}|\rho_\mathrm{opt}) \label{eq:mr fancy pants}
\end{align}
This is convenient because it guarantees that the prescription always reproduces the assumed functional form for $p(\rho_\mathrm{obs}|\theta)$.
We can then choose $p(d | \rho_\mathrm{obs}, \theta)$ to have whichever simple functional form we prefer.\footnote{
In reality, $p(d|\rho_\mathrm{obs},\theta)$ is entirely determined by $p(d,\rho_\mathrm{obs}|\theta)$ and $p(\rho_\mathrm{obs}|\theta)$ and may not be a simple function.
However, one can make simplifying assumptions for $p(d|\rho_\mathrm{obs},\theta)$ and still generate a consistent catalog.
While this may not exactly match the distribution that would be obtained by simulating the full data from first principles, it can nevertheless capture the important correlations within single-event measurement uncertainty.
}

The most straightforward way to do this is to perform a variable transformation, as $\rho_\mathrm{opt}$ and $\theta$ are not independent.
That is, we (re)parametrize each event in terms of $\rho_\mathrm{opt}$ and $\theta_{n-1}$, where $\theta_{n-1}$ denotes the remaining single-event parameters after replacing one with $\rho_\mathrm{opt}$.
We then sample from the posterior over $p(\rho_\mathrm{opt}, \theta_{n-1}|d)$.
Typically, it is convenient to replace the distance to the source with $\rho_\mathrm{opt}$ because the distance is inversely proportional to $\rho_\mathrm{opt}$ given the rest of the single-event parameters.

Our procedure assumes a generative model for a few statistics that determine the shape of the conditioned likelihood $p(d|\rho_\mathrm{obs},\theta_{n-1})$.
We assume a multivariate Normal distribution described by maximum likelihood estimates for $\theta_{n-1}$ (denoted $\hat{\theta}_{n-1}$) and a covariance matrix ($\Sigma$).
Often, the covariance matrix is scaled by $\rho_\mathrm{obs}$ so that events with larger $\rho_\mathrm{obs}$ are more precisely constrained, although this need not be the case.
That is, we assume
\begin{gather}
    p(\hat{\theta}_{n-1}|\Sigma, \theta_{n-1}) = \mathcal{N}(\theta_{n-1}, \Sigma) \\
    \Sigma = \Sigma(\rho_\mathrm{obs})
\end{gather}
so that $\Sigma$ is a deterministic function and the uncertainty in $\theta_{n-1}$ is conditionally independent of the parameter we replaced with $\rho_\mathrm{opt}$.
This is why, for example, we sample in terms of detector-frame masses instead of source-frame masses when replacing the distance with $\rho_\mathrm{opt}$.
It is then straightforward to sample $\theta_{n-1}$ and $\rho_\mathrm{opt}$ from Eq.~\ref{eq:mr fancy pants} with the addition of appropriate priors.
As a final step, we can compute the remaining single-event parameter from $\theta_{n-1}$ and $\rho_\mathrm{opt}$ on a sample-by-sample basis.

In summary, our procedure to generate a synthetic catalog is as follows.
\begin{enumerate}
    \item Sample $(\rho_\mathrm{obs}, \theta) \sim p(\rho_\mathrm{obs}, \theta|\mathbb{D}, \Lambda)$. This is usually done via rejection sampling.
        \begin{enumerate}
            \item Sample $\theta \sim p(\theta|\Lambda)$.
            \item Sample $\rho_\mathrm{obs} \sim p(\rho_\mathrm{obs}|\rho_\mathrm{opt}(\theta))$.
            \item Retain the sample iff $\rho_\mathrm{obs} \geq \mathbb{S}_\mathrm{thr}$.
        \end{enumerate}
    \item Generate the likelihood model.
        \begin{enumerate}
            \item Compute $\Sigma$, which may depend on $\rho_\mathrm{obs}$.
            \item Sample $\hat{\theta}_{n-1} \sim \mathcal{N}(\hat{\theta}_{n-1}, \Sigma)$.
        \end{enumerate}
    \item Generate mock parameter estimation samples with priors $p(\rho_\mathrm{opt})$ and $p(\theta_{n-1})$.
        \begin{enumerate}
            \item Sample $\rho_\mathrm{opt} \sim p(\rho_\mathrm{obs}|\rho_\mathrm{opt}) p(\rho_\mathrm{opt})$.
            \item Sample $\theta_{n-1} \sim \mathcal{N}(\hat{\theta}_{n-1}, \Sigma) p(\theta_{n-1})$.
            \item Compute $\theta_{n}$ (the remaining single-event parameter) from $\theta_{n-1}$ and $\rho_\mathrm{opt}$ on a sample-by-sample basis.
        \end{enumerate}
\end{enumerate}

\citet{Fishbach:2020} implemented a specific example of this in which $\rho_\mathrm{obs}$ is drawn from a Gaussian distribution centered on $\rho_\mathrm{opt}$ with unit variance.\footnote{This is not strictly true for most searches, though~\citep{Essick:2023}.}
The intrinsic parameters (detector-frame masses) are drawn from simple distributions that depend on the true parameters $\theta$ with uncertainties that are scaled by $\rho_\mathrm{obs}$.
Meanwhile, the extrinsic parameters are split into a combination of angles~\citep[that determine the system's orientation;][]{Finn:1993} and the system's distance.
The angle terms are drawn from a simple distribution that does not depend on $\rho_\mathrm{obs}$ or the true parameters.
The distance is fully determined by these angle terms together with $\rho_\mathrm{opt}$ and the other intrinsic parameters.
Finally, source-frame masses are calculated from the sampled detector-frame masses and redshifts.

%------------------------

\subsection{Selection Based on True Event Parameters: $\mathbb{D} \perp d \ | \ \theta$}
\label{sec:parameters}

Alternatively, a relatively common practice is to estimate search sensitivity based on an event's true parameters instead of the observed data.
\begin{equation}
    \mathbb{S}(d, \theta) = \mathbb{S}(\theta)
\end{equation}
A common choice is $\mathbb{S}(\theta) = \rho_\mathrm{opt}(\theta)$.
Note that this must impose some level of inaccuracy in the estimate because real searches do not have access to the event's true parameters.
However, this nevertheless can produce reasonable results because statistics like $\rho_\mathrm{obs}$ are centered on functions of the true parameters, like $\rho_\mathrm{opt}$.
Just as it is simpler to compute and store only $\rho_\mathrm{obs}$ instead of the full data, it is even simpler to only compute $\rho_\mathrm{opt}$.

This choice has repercussions within parameter estimation for individual events, though.
Eq.~\ref{eq:incorrect dag} shows that we should estimate integrals like
\begin{equation} \label{eq:this one}
    \int \diff\theta\, p(d|\theta) p(\theta|\Lambda) \Theta(\mathbb{S}(\theta) \geq \mathbb{S}_\mathrm{thr}),
\end{equation}
whereas we usually evaluate (when $\mathbb{S} = \mathbb{S}(d)$)
\begin{equation}
    \int \diff\theta\, p(d|\theta) p(\theta|\Lambda).
\end{equation}
Conceptually, we can interpret the additional factor $\Theta(\mathbb{S}(\theta) \geq \mathbb{S}_\mathrm{thr})$ as a redefinition of the prior: $q(\theta|\Lambda, \mathbb{S}_\mathrm{thr}) \propto p(\theta|\Lambda) \Theta(\mathbb{S}(\theta) \geq \mathbb{S}_\mathrm{thr})$.
Thus, to be consistent, our estimate for the single-event parameters must incorporate the knowledge that some regions of single-event parameter space are undetectable and are therefore \emph{a priori} excluded.
For example, high redshifts, which correspond to $\rho_\mathrm{opt} < \mathbb{S}_\mathrm{thr}$, should be excluded \textit{a priori} at the level of single-event parameter estimation.
Sec.~\ref{sec:consistency mock pe with selection} shows what can happen when single-event posterior samples are not correctly pruned in this way.
Note that when selection is based on the observed data, noise fluctuations may result in $\rho_\mathrm{obs} \geq \mathbb{S}_\mathrm{thr}$ even for the most distant sources.

Many studies in the literature do not properly include this additional factor when selecting based on the true event parameters.
While this is formally inconsistent, it will matter most for marginally detectable events for which there may naively be significant support \textit{a posteriori} for $\theta$ that correspond to $\mathbb{S}(\theta) < \mathbb{S}_\mathrm{thr}$.
The conclusions of those studies, therefore, may be robust unless they depend sensitively on the behavior of marginally detectable events.
This is one of the reasons why~\citet{Fishbach:2020} developed more realistic selection procedures; they were particularly interested in marginally detectable events at high redshift.
This is also why we primarily saw a bias in the redshift evolution of the merger rate in Sec.~\ref{sec:consistency mock pe with selection} and Sec.~\ref{sec:detected distribution}.

%-------------------------------------------------

\section{Gaussian Toy Models}
\label{sec:gaussian toy models}

We now investigate several toy models to demonstrate that inferences based on Eq.~\ref{eq:incorrect dag rearranged} are inherently biased: they will collapse to the incorrect astrophysical distribution in the limit of infinite events.
This approach has been adopted within the literature for hierarchical inferences of deviations from GR~\citep[see, e.g.,][]{Isi:2019, Isi:2022, GWTC-2-TGR, GWTC-3-TGR} and proposed as a shortcut to avoid estimating selection effects in several GW population analyses~\citep{Rinaldi:2021, Talbot:2023, Sadiq:2022, Sadiq:2023}.

We explore several simplified models to minimize the complexity of the necessary calculations while still showing the intuition described in Sec.~\ref{sec:detected distribution}.
These models should be thought of as existence proofs demonstrating that inferences based on Eq.~\ref{eq:incorrect dag rearranged} cannot be universally consistent alternatives for Eq.~\ref{eq:correct dag}.

%------------------------

\subsection{Analytic Toy Model with Probabilistic Detection}
\label{sec:Gaussian with probabilistic detection}

Consider a simple one-dimensional toy model based on the assumption $(\mathbb{D} \perp \theta \ | \ d)$ chosen so that the relevant integrals remain analytically tractable.\footnote{This is similar to the model considered in~\citet{Essick:2022b}.}
To wit,
\begin{gather}
    p(\theta|\Lambda) = (2\pi \sigma_\Lambda^2)^{-1/2} \exp\left( -\frac{(\theta-\mu_\Lambda)^2}{2\sigma_\Lambda^2}\right) \\
    p(d|\theta) = (2\pi \sigma_o^2)^{-1/2} \exp\left( -\frac{(d-\theta)^2}{2\sigma_o^2}\right) \\
    P(\mathbb{D}|d) = \exp\left( -\frac{(d-\mu_D)^2}{2\sigma_D^2}\right)
\end{gather}
Within this model, we obtain
\begin{gather}
    p(d|\Lambda) = \left(2\pi (\sigma_\Lambda^2 + \sigma_o^2) \right)^{-1/2} \exp\left( -\frac{(d-\mu_\Lambda)^2}{2(\sigma_\Lambda^2 + \sigma_o^2)} \right) \\
    P(\mathbb{D}|\Lambda) = \left( \frac{\sigma_D^2}{\sigma_\Lambda^2 + \sigma_o^2 + \sigma_D^2} \right) \exp\left( -\frac{(\mu_D - \mu_\Lambda)^2}{2(\sigma_\Lambda^2 + \sigma_o^2 + \sigma_D^2)} \right)
\end{gather}
so that Eq.~\ref{eq:correct dag} becomes
%\begin{align}
%    p(\Lambda| & \{d_i,\mathbb{D}_i\}, N) \propto \nonumber \\
%      & \left(\frac{\sigma_\Lambda^2 + \sigma_o^2 + \sigma_D^2}{2\pi \sigma_D^2 (\sigma_\Lambda^2 + \sigma_o^2)}\right)^{N/2} \nonumber \\
%      & \quad\quad \times \exp\left( -\frac{\sum_i^N (d_i-\mu_\Lambda)^2}{2(\sigma_\Lambda^2 + \sigma_o^2)} + \frac{N(\mu_D - \mu_\Lambda)^2}{2(\sigma_\Lambda^2 + \sigma_o^2 + \sigma_D^2)} \right) \label{eq:i like this eq}
%\end{align}
\begin{equation}
    p(\Lambda| \{d_i,\mathbb{D}_i\}, N) \propto
      \left(\frac{\sigma_\Lambda^2 + \sigma_o^2 + \sigma_D^2}{2\pi \sigma_D^2 (\sigma_\Lambda^2 + \sigma_o^2)}\right)^{N/2}
      \exp\left( -\frac{\sum_i^N (d_i-\mu_\Lambda)^2}{2(\sigma_\Lambda^2 + \sigma_o^2)} + \frac{N(\mu_D - \mu_\Lambda)^2}{2(\sigma_\Lambda^2 + \sigma_o^2 + \sigma_D^2)} \right) \label{eq:i like this eq}
\end{equation}
assuming a flat prior $p(\Lambda)$.
From this, we can define estimators by maximizing Eq.~\ref{eq:i like this eq} with respect to $\mu_\Lambda$ and $\sigma_\Lambda^2$:
\begin{align}
   \hat{\mu}_\Lambda & = \frac{(\sigma_\Lambda^2 + \sigma_o^2 + \sigma_D^2) m_d - (\sigma_\Lambda^2 + \sigma_o^2) \mu_D}{\sigma_D^2} \label{eq:hat mu lambda} \\
   \hat{\sigma}^2_\Lambda & = \frac{\sigma_D^2 (m_{d^2} - m_d^2)}{\sigma_d^2 - (m_{d^2} - m_d^2)} - \sigma_o^2 \label{eq:hat sigma lambda}
\end{align}
where we define the moments of the observed data
\begin{align}
    m_d & = \frac{1}{N}\sum\limits_i^N d_i \\
    m_{d^2} & = \frac{1}{N}\sum\limits_i^N d_i^2
\end{align}
A few other distributions (conditioned on detection and the population) will also be useful references.
\begin{widetext}
\begin{align}
    p(\theta|\mathbb{D},\Lambda) & = \left( \frac{\sigma_\Lambda^2 + \sigma_o^2 + \sigma_D^2}{2\pi \sigma_\Lambda^2 (\sigma_o^2 + \sigma_D^2)} \right)^{1/2} \exp\left( -\frac{\sigma_\Lambda^2 + \sigma_o^2 + \sigma_D^2}{2\sigma_\Lambda^2 (\sigma_o^2 + \sigma_D^2)} \left( \theta - \frac{\sigma_\Lambda^2 \mu_D + (\sigma_o^2 + \sigma_D^2)\mu_\Lambda}{\sigma_\Lambda^2 + \sigma_o^2 + \sigma_D^2} \right)^2 \right) \label{eq:gaussian3 good theta given det} \\
    p(d|\mathbb{D},\Lambda) & = \left( \frac{\sigma_\Lambda^2 + \sigma_o^2 + \sigma_D^2}{2\pi(\sigma_\Lambda^2 + \sigma_o^2)\sigma_D^2}\right)^{1/2} \exp\left( -\frac{\sigma_\Lambda^2 + \sigma_o^2 + \sigma_D^2}{2(\sigma_\Lambda^2 + \sigma_o^2)\sigma_D^2} \left( d - \frac{\sigma_D^2 \mu_\Lambda + (\sigma_\Lambda^2 + \sigma_o^2)\mu_D}{\sigma_\Lambda^2 + \sigma_o^2 + \sigma_D^2} \right)^2 \right) \label{eq:gaussian3 good data given det}
\end{align}
\end{widetext}
In particular, Eq.~\ref{eq:gaussian3 good data given det} shows that the moments of the detected data will approach
\begin{align}
    \lim\limits_{N\rightarrow\infty} m_d & = \frac{\sigma_D^2 \mu_\Lambda + (\sigma_\Lambda^2 + \sigma_o^2)\mu_D}{\sigma_\Lambda^2 + \sigma_o^2 + \sigma_D^2} \label{eq:gaussian2 data m1} \\
    \lim\limits_{N\rightarrow\infty} \left( m_{d^2} - m_d^2 \right) & = \frac{(\sigma_\Lambda^2 + \sigma_o^2)\sigma_D^2}{\sigma_\Lambda^2 + \sigma_o^2 + \sigma_D^2} \label{eq:gaussian2 data m2}
\end{align}
Inserting Eqs.~\ref{eq:gaussian2 data m1} and~\ref{eq:gaussian2 data m2} into Eqs.~\ref{eq:hat mu lambda} and~\ref{eq:hat sigma lambda}, we (unsurprisingly) obtain
\begin{align}
    \lim\limits_{N\rightarrow\infty} \hat{\mu}_\Lambda & = \mu_\Lambda \\
    \lim\limits_{N\rightarrow\infty} \hat{\sigma}^2_\Lambda & = \sigma^2_\Lambda
\end{align}
demonstrating that the inference is unbiased.

Now, consider the toy model where we instead first infer $q(\theta|\mathbb{D},\Lambda)$ from Eq.~\ref{eq:incorrect dag rearranged} and then divide by $P(\mathbb{D}|\theta)$ to obtain $q(\theta|\Lambda)$.
Because all the distributions are Gaussian, we adopt a Gaussian ansatz.\footnote{This toy model is similar to what was used to motivate hierarchical inference within tests of GR in~\citet{Isi:2019, Isi:2022}.}
\begin{equation}
    q(\theta|\mathbb{D},\Lambda) = (2\pi\sigma^2)^{-1/2} \exp\left( -\frac{(\theta-\mu)^2}{2\sigma^2}\right)
\end{equation}
The hyperposterior under this model is
%\begin{multline}
%    q(\Lambda|\{d_i,D_i\}, N) \propto \\ (2\pi (\sigma^2 + \sigma_o^2))^{-N/2} \exp\left( -\frac{\sum_i^N (d_i-\mu)^2}{2(\sigma^2 + \sigma_o^2)} \right)
%\end{multline}
\begin{equation}
    q(\Lambda|\{d_i,D_i\}, N) \propto (2\pi (\sigma^2 + \sigma_o^2))^{-N/2} \exp\left( -\frac{\sum_i^N (d_i-\mu)^2}{2(\sigma^2 + \sigma_o^2)} \right)
\end{equation}
and the associated estimators are
\begin{align}
    \hat{\mu} & = m_d \label{eq:gaussian3 bad mu} \\
    \hat{\sigma}^2 & = (m_{d^2} - m_d^2) - \sigma_o^2 \label{eq:gaussian3 bad sigma}
\end{align}
Now, assuming we can measure $P(\mathbb{D}|\theta)$ perfectly, we will only infer the correct astrophysical distribution $p(\theta|\Lambda)$ if we infer $q(\theta|\mathbb{D},\Lambda) = p(\theta|\mathbb{D},\Lambda)$ correctly.
Comparing Eqs.~\ref{eq:gaussian3 bad mu} and~\ref{eq:gaussian3 bad sigma} with the moments from Eq.~\ref{eq:gaussian3 good theta given det}, we find
\begin{align}
    \lim\limits_{N\rightarrow\infty} \left( \hat{\mu} - \mathrm{E}[\theta]_{p(\theta|\mathbb{D},\Lambda)} \right) & = \frac{\sigma_o^2}{\sigma_\Lambda^2 + \sigma_o^2 + \sigma_D^2} (\mu_D - \mu_\Lambda) \label{eq:bias in mean} \\
    \lim\limits_{N\rightarrow\infty} \left( \hat{\sigma}^2 - \mathrm{V}[\theta]_{p(\theta|\mathbb{D},\Lambda)} \right) & = - \frac{(2\sigma_\Lambda^2 + \sigma_o^2) \sigma_o^2}{\sigma_\Lambda^2 + \sigma_o^2 + \sigma_D^2} < 0
\end{align}
and conclude that such an approach is biased.
That is, in the limit of an infinite number of events, an inference based on Eq.~\ref{eq:incorrect dag rearranged} will not collapse to the correct astrophysical distribution $p(\theta|\Lambda)$ because it does not recover the correct distribution of the true parameters of detected events: $q(\theta|\mathbb{D},\Lambda) \neq p(\theta|\mathbb{D},\Lambda)$.
Note that this is the case even though the incorrect inference can exactly reproduce the properties of the distribution of detected data: $q(d|\mathbb{D},\Lambda) = p(d|\mathbb{D},\Lambda)$.

The incorrect inference may not always be able to reproduce the distribution of detected data, though.
In particular, if $\sigma_D \ll \sigma_o$, then $(m_{d^2} - m_d^2) \sim \sigma_D^2$ and $\hat{\sigma}^2 < 0$.
Similarly, if $\sigma_\Lambda \ll \sigma_o,\, \sigma_D$, then we also obtain $\hat{\sigma}^2 < 0$.
This is unphysical, and the incorrect inference will over-predict the variance in the detected data even as it collapses to a delta-function $q(\theta|\mathbb{D},\Lambda) \sim \delta(\theta - \hat{\mu})$.
As such, a test of GR based on Eq.~\ref{eq:incorrect dag rearranged} that appears to confidently confirm GR's prediction (inferred distributions are $\delta$-functions centered on GR~\citep{Isi:2019, Isi:2022}) does not actually mean that GR is the correct physical theory of gravity, particularly if searches are strongly tuned to only detect signals that resemble GR (small $\sigma_D$).
What's more, a test of GR based on Eq.~\ref{eq:incorrect dag rearranged} that infers a delta-function that is \emph{not} centered on GR may not collapse to the correct value of the deviation.
Eq.~\ref{eq:bias in mean} shows there can still be a bias when $\sigma_\Lambda = 0$ if the detection is not centered on the true parameter ($\mu_D \neq \mu_\Lambda$).

Although this is a simplistic toy model, we again stress that the failure of Eq.~\ref{eq:incorrect dag rearranged} in at least one case shows that it cannot be a universally self-consistent alternative to the standard approach (Eq.~\ref{eq:correct dag}).

It is enlightening to examine the limits in which the bias can be made small.
First, considering the bias in $\hat{\mu}$, we see that it is small when any of the following are true.
\begin{itemize}
    \item $\sigma_D \gg \sigma_\Lambda, \sigma_o$ : everything is detectable (i.e., $\mathbb{D} \perp \theta, d$).
    \item $\sigma_o \ll \sigma_\Lambda, \sigma_D$ : perfect measurements (i.e., $d$ is one-to-one with $\theta$).
    \item $\sigma_\Lambda \gg \sigma_o, \sigma_D$ : the population is extremely flat (and $\mu_\Lambda$ is meaningless).
    \item $\mu_D = \mu_\Lambda$ : a fine-tuned solution in which the detection probability just happens to be centered on the population.
\end{itemize}
Considering the bias in $\hat{\sigma}^2$, we find that it is small only when at least one of the first two conditions is met.
What's more, these are exactly the conditions that render the assumptions $(\mathbb{D} \perp d \ | \ \theta)$ and $(\mathbb{D} \perp \theta \ | \ d)$ compatible (see Sec.~\ref{sec:graphs}).

%------------------------

\subsection{Toy Model with Deterministic Detection}
\label{sec:Gaussian with deterministic detection}

As shown in Sec.~\ref{sec:graphs}, the correct model does not care about whether selection is probabilistic or deterministic, and Sec.~\ref{sec:Gaussian with probabilistic detection} provides an explicit example with probabilistic detection.
For completeness, we also present an example with deterministic selection, similar to the models investigated in Appendix~\ref{sec:consistent pe and injections}.

We assume the same toy model as Sec.~\ref{sec:Gaussian with probabilistic detection} but adopt a deterministic selection (i.e., the data are censored below $d_\mathrm{min}$).
\begin{equation}
    P(\mathbb{D}|d) = \Theta(d_\mathrm{min} \leq d)
\end{equation}
It is trivial to show that the distribution of detected data is a truncated Gaussian.
Importantly, there is no prior $q(\theta|\mathbb{D},\Lambda)$ that can correctly reproduce a truncated Gaussian for $q(d|\mathbb{D},\Lambda)$ within Eq.~\ref{eq:incorrect dag rearranged} when the likelihood is a Gaussian $p(d|\theta) = \mathcal{N}(\theta, \sigma_o^2)$.
As such, from the start, we see that the incorrect assumption within Eq.~\ref{eq:incorrect dag rearranged} introduces model misspecification; it will never reproduce the correct distribution of detected data.
We should not expect it to infer the correct astrophysical distribution, either.

\begin{figure*}
    \begin{minipage}{0.33\textwidth}
        \begin{center}
            \Large{$10^2$ events} \\
            \vspace{0.25cm}
            \includegraphics[width=1.0\textwidth]{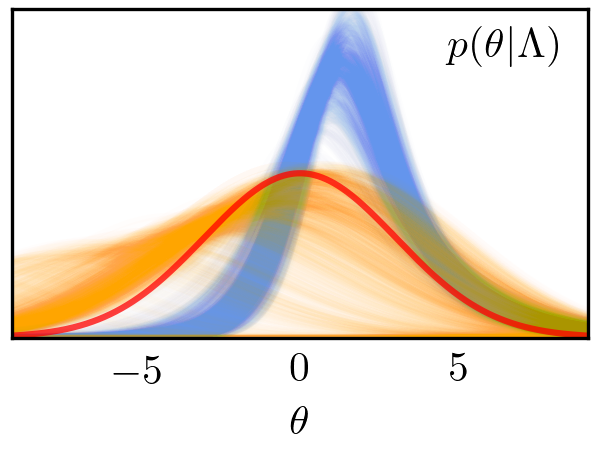} \\
            \includegraphics[width=1.0\textwidth]{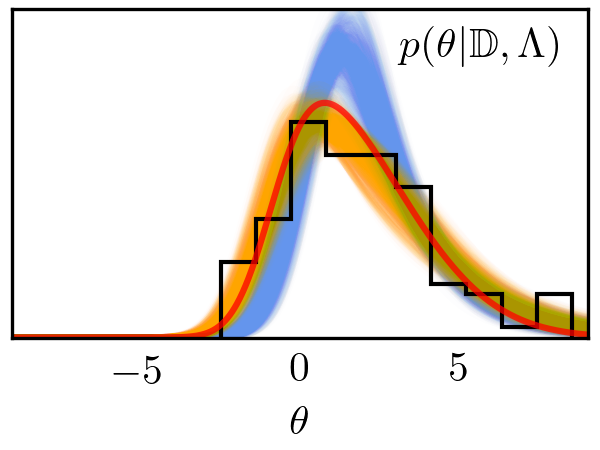} \\
            \includegraphics[width=1.0\textwidth]{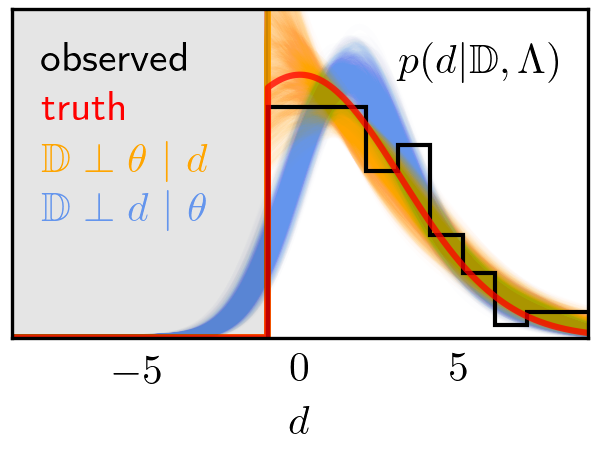}
        \end{center}
    \end{minipage}
    \begin{minipage}{0.33\textwidth}
        \begin{center}
            \Large{$10^3$ events} \\
            \vspace{0.25cm}
            \includegraphics[width=1.0\textwidth]{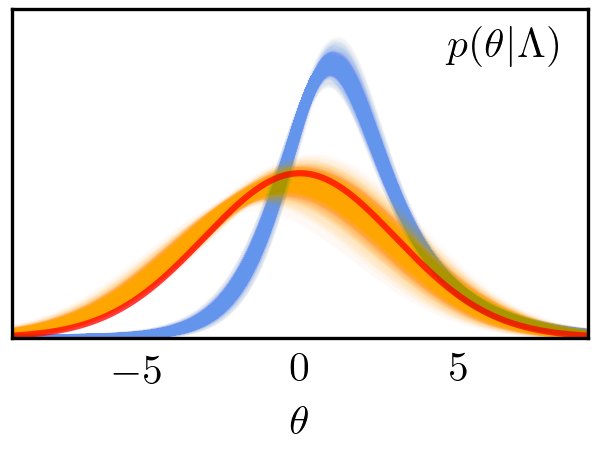} \\
            \includegraphics[width=1.0\textwidth]{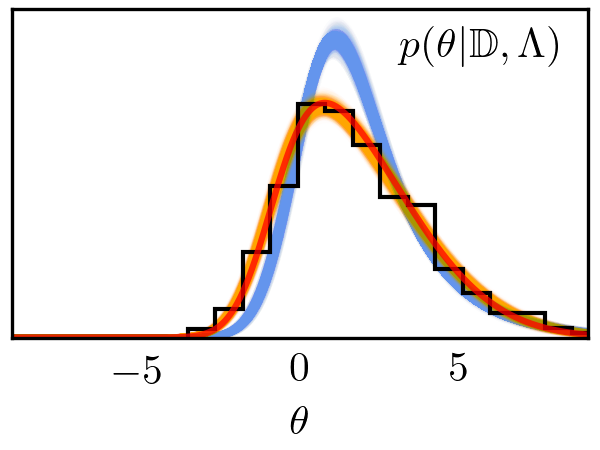} \\
            \includegraphics[width=1.0\textwidth]{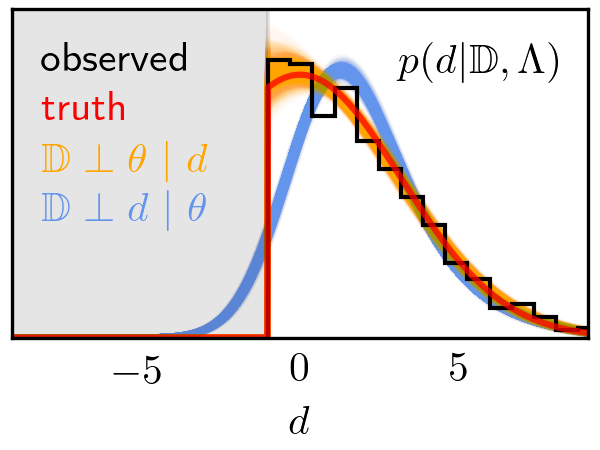}
        \end{center}
    \end{minipage}
    \begin{minipage}{0.33\textwidth}
        \begin{center}
            \Large{$10^4$ events} \\
            \vspace{0.25cm}
            \includegraphics[width=1.0\textwidth]{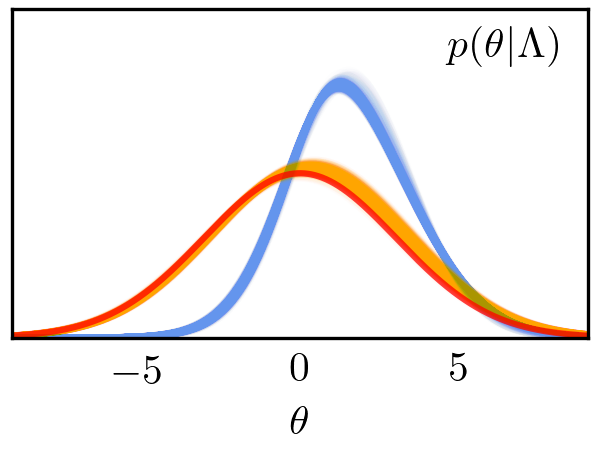} \\
            \includegraphics[width=1.0\textwidth]{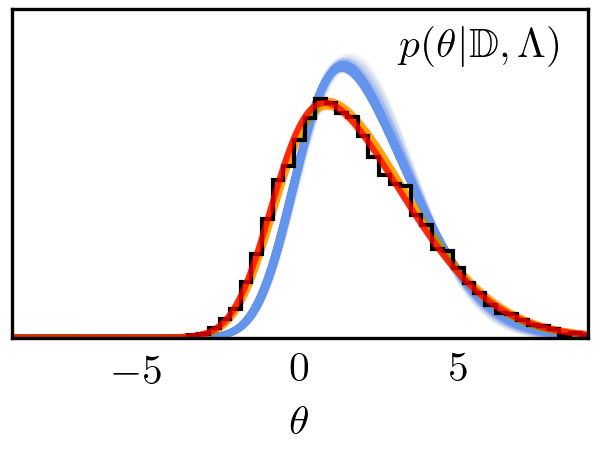} \\
            \includegraphics[width=1.0\textwidth]{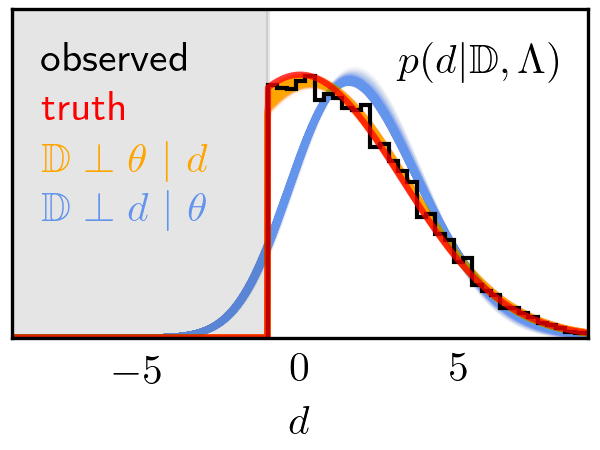}
        \end{center}
    \end{minipage}
    \caption{
        Inferred distributions with deterministic selection.
        (\emph{left to right}) Catalogs with $10^2$, $10^3$, and $10^4$ events generated with $\mu_\Lambda = 0$, $\sigma_\Lambda = 3$, $\sigma_o = 1$, and $d_\mathrm{min} = -1$.
        (\emph{top to bottom}) The astrophysical distribution $p(\theta|\Lambda)$, the distribution of true parameters for detected events $p(\theta|\mathbb{D},\Lambda)$, and the distribution of detected data $p(d|\mathbb{D},\Lambda)$ with (\emph{shaded}) the censored region ($d < d_\mathrm{min}$).
        We show (\emph{black}) histograms of the know parameters of events within the simulated catalogs, (\emph{red}) the true distribution, ($\emph{orange}$) the inferred distribution assuming the physical DAG, and (\emph{blue}) the inferred distribution assuming the unphysical DAG (i.e., fitting for the ``detected distribution'' $q(\theta|\mathbb{D},\Lambda)$ and dividing by selection effects $P(\mathbb{D}|\theta)$ \textit{post hoc}).
        For the inferred distributions, each line represents a single draw from the corresponding hyperposterior.
    }
    \label{fig:Gaussian with deterministic detection}
\end{figure*}

Nevertheless, we attempt to minimize the model misspecification by adopting an ansatz for $q(\theta|\mathbb{D},\Lambda)$ that matches the known functional form from the physical DAG.
\begin{equation}
    q(\theta|\mathbb{D},\Lambda) = \frac{p(\theta|\Lambda) \int\limits_{d_\mathrm{min}}^\infty \diff d\, p(d|\theta)}{P(\mathbb{D}|\Lambda)}
\end{equation}
where
\begin{equation}
    P(\mathbb{D}|\Lambda) = \int \diff\theta\, p(\theta|\Lambda) \int\limits_{d_\mathrm{min}}^\infty \diff d\, p(d|\theta)
\end{equation}
We sample from the hyperposterior for both the correct (Eq.~\ref{eq:correct dag}) and incorrect (Eq.~\ref{eq:incorrect dag rearranged}) inferences.
Fig.~\ref{fig:Gaussian with deterministic detection} compares the various inferred distributions.
As expected, we find that the physical DAG is able to correctly infer all the distributions and collapses to a more precise estimate as the catalog size increases.
On the contrary, the unphysical DAG does not correctly infer any of the distributions and collapses to an equally precise but biased result as the size of the catalog grows.
Specifically, we see that the unphysical inference produces $q(d|\mathbb{D},\Lambda)$ that is too wide along with $q(\theta|\mathbb{D},\Lambda)$ that is too narrow in analogy to the well known ``edge effects'' in kernel density estimates like those used in~\citet{Sadiq:2022, Sadiq:2023}.
This forces the estimate for $q(\theta|\Lambda)$ obtained from Eq.~\ref{eq:divide by selection} assuming $p(\theta|\mathbb{D},\Lambda) = q(\theta|\mathbb{D},\Lambda)$ to also be too narrow.

Interestingly, we may be able to infer which DAG from Fig.~\ref{fig:simplified dags} is correct directly from the data in the presence of deterministic selection functions, as the incorrect DAG predicts a much worse fit to the observed data than the physical DAG.
Indeed, this is already apparent within real analyses; see discussion in ~\citet{Essick:2023}.

%-------------------------------------------------

\bibliography{refs.bib}

%-------------------------------------------------
\end{document}